\documentclass{aa}
\usepackage{graphicx}
\usepackage{txfonts}

\usepackage{subfig}
\usepackage{float}
\usepackage{comment}
\usepackage{caption}
\usepackage{bm}
\usepackage{amsfonts}
\usepackage{amssymb}
\usepackage{array}
\usepackage{natbib}
\usepackage{tikz}
\usepackage{booktabs}
\newcommand{\argmin}{\mathop{\mathrm{arg\:min}}}

\usepackage{url}

\usepackage{color}
\usepackage{multirow}

\usepackage[ruled,vlined]{algorithm2e}

\begin{document}

\title{Imaging from STIX visibility amplitudes}

\author{Paolo Massa\inst{1} \and Emma Perracchione\inst{2} \and Sara Garbarino\inst{1} \and Andrea F. Battaglia\inst{4,5} \and Federico Benvenuto\inst{1}  \and Michele Piana\inst{1,2} \and Gordon Hurford\inst{4}  \and S\"am Krucker\inst{3,4}}
\institute{MIDA, Dipartimento di Matematica,  Universit\`{a} degli Studi di Genova, Via   Dodecaneso 35, 16146 Genova, Italy\\
\and
CNR-SPIN, Via Dodecaneso 33, 16146 Genova, Italy\\
\and
Space Sciences Laboratory, University of California, 7 Gauss Way, 94720 Berkeley, USA\\
\and
University of Applied Sciences and Arts Northwestern Switzerland, Bahnhofstrasse 6, 5210 Windisch, Switzerland  \\ 
\and
Swiss Federal Institute of Technology in Zurich (ETHZ), R\"amistrasse 101, 8039 Zurich, Switzerland \\
\email{massa.p@dima.unige.it},
\email{perracchione@dima.unige.it},  \email{garbarino@dima.unige.it},
\email{benvenuto@dima.unige.it},
\email{piana@dima.unige.it},
\email{gordon.hurford@fhnw.ch},
\email{andrea.battaglia@fhnw.ch}, \email{krucker@berkeley.edu}}


\abstract
{}
{To provide the first demonstration of STIX Fourier-transform X-ray imaging using semi-calibrated (amplitude-only) visibility data acquired during the Solar Orbiter's cruise phase.}
{We use a parametric imaging approach by which STIX visibility amplitudes are fitted by means of two non-linear optimization methods: a fast meta-heuristic technique inspired by social behavior, and a Bayesian Monte Carlo sampling method, which, although slower, provides better quantification of uncertainties. }
{When applied to a set of solar flare visibility amplitudes recorded by STIX on November 18, 2020 the two parametric methods provide very coherent results. The analysis also demonstrates the ability of STIX to reconstruct high time resolution information and, from a spectral viewpoint, shows the reliability of a double-source scenario consistent with a thermal versus nonthermal interpretation. }
{In this preliminary analysis of STIX imaging based only on visibility amplitudes, we formulate the imaging problem as a non-linear parametric issue we addressed by means of two high-performance optimization techniques that both showed the ability to sample the parametric space in an effective fashion, thus avoiding local minima.}


\titlerunning{Imaging from STIX visibility amplitudes}
\authorrunning{Massa et al}

\maketitle

\date{\today}

\section{Introduction}

The solar corona is permanently heated to temperatures of several million degrees, well above the moderately hot solar surface which is around 6000 K. While the source of the energy that heats the corona has been early on identified as the solar magnetic field, the detailed process(es) of how the energy release is triggered and how energy is eventually dissipated into heat has been an ongoing research topic for many decades. The recently launched missions Parker Solar Probe and Solar Orbiter open new windows to unravel this mystery.  Hard X-ray observations provide strong diagnostics of the hottest solar plasma and  nonthermal electrons through the bremsstrahlung process, and therefore they play a key role in investigating the impulsive magnetic energy release in the corona during solar flares. While hard X-ray focusing optics has made great progress in the past years \citep[e.g.,][]{2013SPIE.8862E..0RK}, indirect imaging systems in hard X-rays have had great success with instruments such as the {{Hard X-ray Telescope}} on-board {{Yohkoh}} \citep{1992PASJ...44L..45K} and the {{Reuven Ramaty High Energy Solar Spectroscopic Imager (RHESSI)}} \citep{2002SoPh..210....3L}. 

The {{Spectrometer/Telescope for Imaging X-rays (STIX)}} \citep{krucker2020spectrometer} is a hard X-ray imaging-spectrometer onboard the Solar Orbiter spacecraft.  STIX imaging uses an indirect Fourier technique in which the native form of the data is a set of angular Fourier components. Imaging-spectroscopy is based on the choice of the energy of the photons used as the input to the imaging process. The hardware includes a set of $30$ independent subcollimators, each consisting of a coarsely-pixelated photon detector with good energy resolution located behind a pair of widely separated grids whose joint transmission creates a large-scale Moiré pattern.  Each Moiré pattern can be interpreted to measure a Fourier component of the incoming flux, termed a {\em{visibility}}, whose amplitude corresponds to the difference between the maximum and the minimum of the Moiré pattern and whose phase corresponds to the location of the peak of the Moiré pattern \citep{STIX1,2019A&A...624A.130M}.  The angular frequency, which corresponds to each such visibility, is fixed by the choice of orientation and pitch of the corresponding grids.  The choices can be divided into 10 groups, each of which corresponds to one of 10 logarithmically-spaced angular resolutions in the range 7.1 to 180 arcseconds as measured at 3 different orientations. 

For a fully-calibrated instrument, the image reconstruction problem for STIX is therefore the linear Fourier inversion problem from limited data that can be addressed by means of regularization methods like interpolation/extrapolation \citep{2020arXiv201214007P}, maximum entropy \citep{massa2020mem_ge}, or compressed sensing \citep{duval2018solar}. 

However, while the current calibration of the STIX imaging system is reliable for visibility amplitudes,  the more complex phase calibration is just not yet available (while we are writing this paper, the calibration and validation process for STIX visibilities is in progress and will be finalized well in advance with respect to the beginning of the nominal phase of the mission in September 2021). Therefore, the current image reconstruction problem is the one of determining (partial) information on the flaring source from measurements of the visibility amplitudes only.   Such a reconstruction problem has two main difficulties: 1) it is non--linear, and 2) no information on the source location can be retrieved from the amplitude of the Fourier transform alone. We point out that this situation is analogous to amplitude-only imaging in the early days of very long baseline radio interferometry \citep{1984ARA&A..22...97P} as well as in other fields.

This study describes an approach to the solution of the STIX image reconstruction problem from visibility amplitudes, in which the flaring source is modelled by means of a limited number of simple but flare--appropriate parametric shapes \citep{2009ApJ...698.2131D}. We address the problem of source parameter estimation with two alternative methods. The first one is based on a stochastic optimization technique \citep{eberhart1995particle} for solving the forward--fitting problem. It also relies on a confidence strip approach \citep{1994A&A...288..949P} for computing parameter uncertainties, which has the advantage of being fast, but comes at the price of having to perturb multiple times the input data, often resulting in misinterpretations. The second one is a Bayesian technique \citep{sciacchitano2019sparse}, which produces full probabilistic description of the source parameters in a mathematically sound way, and often produces smaller uncertainties, but has higher computational cost. We point out that forward-fitting techniques have been already used within the framework of Fourier imagers missions. For example, a forward-fitting algorithm based on deterministic optimization is in the Solar Software (SSW) tree of the RHESSI mission \citep{aschwanden2003reconstruction} and the Bayesian technique illustrated in \cite{sciacchitano2018identification} has been validated against fully-calibrated RHESSI visibilities.

We illustrate the two parametric methods using STIX visibility amplitudes associated to a flaring event of the Sun in November 2020. The images provided by this example should not be considered as science products. However, this analysis may represent a timely demonstration of STIX imaging capabilities in terms of temporal and spectral resolution. 

The plan of the paper is as follows. Section 2 introduces the visibility amplitude imaging problem. Section 3 describes the computational approach to its solution and Section 4 contains the results of the application of such approach to a set of experimental STIX observations. Our conclusions are offered in Section 5.

\section{The visibility amplitude imaging problem}

\begin{figure*}[h]
\begin{center}

\tikzset{every picture/.style={line width=0.75pt}} 

\begin{tikzpicture}[x=0.6pt,y=0.6pt,yscale=-1,xscale=1]

\draw    (100.33,201.67) -- (99.68,42.67) ;
\draw [shift={(99.67,39.67)}, rotate = 449.76] [fill={rgb, 255:red, 0; green, 0; blue, 0 }  ][line width=0.08]  [draw opacity=0] (8.93,-4.29) -- (0,0) -- (8.93,4.29) -- cycle    ;
\draw [line width=0.75]    (20.67,120.83) -- (176.33,120.51) ;
\draw [shift={(179.33,120.5)}, rotate = 539.88] [fill={rgb, 255:red, 0; green, 0; blue, 0 }  ][line width=0.08]  [draw opacity=0] (8.93,-4.29) -- (0,0) -- (8.93,4.29) -- cycle    ;
\draw  [color={rgb, 255:red, 0; green, 24; blue, 255 }  ,draw opacity=1 ][line width=1.5]  (60.87,120.67) .. controls (60.87,98.69) and (78.39,80.88) .. (100,80.88) .. controls (121.61,80.88) and (139.13,98.69) .. (139.13,120.67) .. controls (139.13,142.64) and (121.61,160.46) .. (100,160.46) .. controls (78.39,160.46) and (60.87,142.64) .. (60.87,120.67) -- cycle ;
\draw  [dash pattern={on 4.5pt off 4.5pt}]  (69.67,94.67) -- (130.33,146.67) ;
\draw    (310.33,201) -- (309.68,42) ;
\draw [shift={(309.67,39)}, rotate = 449.76] [fill={rgb, 255:red, 0; green, 0; blue, 0 }  ][line width=0.08]  [draw opacity=0] (8.93,-4.29) -- (0,0) -- (8.93,4.29) -- cycle    ;
\draw [line width=0.75]    (230.67,120.17) -- (386.33,119.84) ;
\draw [shift={(389.33,119.83)}, rotate = 539.88] [fill={rgb, 255:red, 0; green, 0; blue, 0 }  ][line width=0.08]  [draw opacity=0] (8.93,-4.29) -- (0,0) -- (8.93,4.29) -- cycle    ;
\draw    (520.33,201) -- (519.68,42) ;
\draw [shift={(519.67,39)}, rotate = 449.76] [fill={rgb, 255:red, 0; green, 0; blue, 0 }  ][line width=0.08]  [draw opacity=0] (8.93,-4.29) -- (0,0) -- (8.93,4.29) -- cycle    ;
\draw [line width=0.75]    (440.67,120.17) -- (596.33,119.84) ;
\draw [shift={(599.33,119.83)}, rotate = 539.88] [fill={rgb, 255:red, 0; green, 0; blue, 0 }  ][line width=0.08]  [draw opacity=0] (8.93,-4.29) -- (0,0) -- (8.93,4.29) -- cycle    ;
\draw  [color={rgb, 255:red, 0; green, 24; blue, 255 }  ,draw opacity=1 ][line width=1.5]  (457.77,77) .. controls (457.77,66.01) and (466.53,57.1) .. (477.33,57.1) .. controls (488.14,57.1) and (496.9,66.01) .. (496.9,77) .. controls (496.9,87.99) and (488.14,96.9) .. (477.33,96.9) .. controls (466.53,96.9) and (457.77,87.99) .. (457.77,77) -- cycle ;
\draw  [color={rgb, 255:red, 0; green, 24; blue, 255 }  ,draw opacity=1 ][line width=1.5]  (273.18,159.04) .. controls (262.27,148.75) and (269.9,122.93) .. (290.24,101.36) .. controls (310.57,79.8) and (335.9,70.66) .. (346.82,80.96) .. controls (357.73,91.25) and (350.1,117.07) .. (329.76,138.64) .. controls (309.43,160.2) and (284.1,169.34) .. (273.18,159.04) -- cycle ;
\draw  [dash pattern={on 4.5pt off 4.5pt}]  (291.95,99.51) -- (328.05,140.49) ;
\draw  [dash pattern={on 4.5pt off 4.5pt}]  (273.18,159.04) -- (346.82,80.96) ;
\draw  [draw opacity=0] (322.79,105.97) .. controls (326.7,109.42) and (329.17,114.43) .. (329.17,120) .. controls (329.17,120.11) and (329.17,120.22) .. (329.16,120.32) -- (310,120) -- cycle ; \draw   (322.79,105.97) .. controls (326.7,109.42) and (329.17,114.43) .. (329.17,120) .. controls (329.17,120.11) and (329.17,120.22) .. (329.16,120.32) ;
\draw  [color={rgb, 255:red, 0; green, 24; blue, 255 }  ,draw opacity=1 ][line width=1.5]  (533.94,163) .. controls (533.94,147.78) and (546.8,135.44) .. (562.67,135.44) .. controls (578.53,135.44) and (591.39,147.78) .. (591.39,163) .. controls (591.39,178.22) and (578.53,190.56) .. (562.67,190.56) .. controls (546.8,190.56) and (533.94,178.22) .. (533.94,163) -- cycle ;
\draw  [dash pattern={on 4.5pt off 4.5pt}]  (463.67,64) -- (491,90) ;
\draw  [dash pattern={on 4.5pt off 4.5pt}]  (541.67,144.83) -- (583.67,181.17) ;
\draw  [dash pattern={on 0.84pt off 2.51pt}]  (562.67,163) -- (520.33,163) ;
\draw  [dash pattern={on 0.84pt off 2.51pt}]  (562.67,163) -- (562.33,119.67) ;

\draw (132.33,150.07) node [anchor=north west][inner sep=0.75pt]    {$f$};
\draw (266.05,77.89) node [anchor=north west][inner sep=0.75pt]    {$f_{m}$};
\draw (247.38,161.22) node [anchor=north west][inner sep=0.75pt]    {$f_{M}$};
\draw (330,103.) node [anchor=north west][inner sep=0.75pt]    {$\alpha $};
\draw (440.05,43.22) node [anchor=north west][inner sep=0.75pt]    {$f_{1}$};
\draw (584.05,177.89) node [anchor=north west][inner sep=0.75pt]    {$f_{2}$};
\draw (554,100.73) node [anchor=north west][inner sep=0.75pt]    {$x_{0}$};
\draw (501.33,153.4) node [anchor=north west][inner sep=0.75pt]    {$y_{0}$};

\end{tikzpicture}

\end{center}
\caption{Gaussian shapes considered in the parametric imaging process.}\label{fig:fig-1}
\end{figure*}
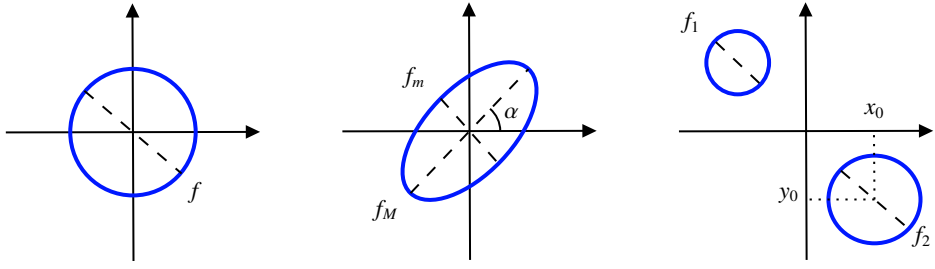

When a stable and calibrated imaging pipeline will become available, the STIX imaging problem will be described by

\begin{equation}\label{b1}
{\mathbf{V}} = {\mathbf{F}}\varphi~,
\end{equation}
where $\varphi=\varphi({\mathbf{x}})$ is the incoming photon flux emitted from location ${\mathbf{x}}=(x,y)$ on the solar disk, ${\mathbf{V}}$ is the set of $N$ visibilities measured by the telescope in correspondence of the $N$ points $\{(u_j,v_j)\}_{j=1}^{N}$ on the $(u,v)$-plane made of all angular frequencies, and ${\mathbf{F}}$ is the imaging operator that maps the functions representing the flaring source into samples of their Fourier transform, i.e.
\begin{equation}\label{b2}
({\bf{F}}\varphi)_j = \int\int dx dy \varphi(x,y) e^{i2\pi(u_j x + v_j y)}~~~j=1,\ldots,N~.
\end{equation}
The $30$ STIX subcollimators will provide $N=30$ visibilities. We note that the STIX imaging problem, i.e. the one of reconstructing the photon flux from $N$ visibilities, is linear, but does not have a unique solution.

Unfortunately, the currently available calibration of the imaging system is limited to the visibility amplitudes, the phase calibration being more complex and therefore still under construction. The formal consequence of this limitation is that the STIX imaging operator at this stage is represented by
\begin{equation}\label{b3}
({\mathbf{F}}_a (\varphi))_j = \left|\int\int dx dy \varphi(x,y) e^{i2\pi(u_j x + v_j y)} \right|~~~j=1,\ldots,N~,
\end{equation}
and therefore the imaging problem addressed in the present study is
\begin{equation}\label{b4}
{\mathbf{A}} = {\mathbf{F}}_a(\varphi)~,
\end{equation}
where $\mathbf{A} = |{\bf V}|$ and  $| \cdot|$ should be intended as component-wise. This imaging problem is more challenging to solve than (\ref{b1}) for two key reasons: 1) it is non-linear, and 2) visibility amplitudes do not encode any information on the position of the flaring source. To overcome these limitations, we implemented forward--fitting procedures to estimate the parameters of three different parametric shapes of very simple architecture.
Specifically, we have considered:
\begin{itemize}
\item A Gaussian circular source
\begin{equation}
\varphi_C(x,y) = \frac{\phi}{2 \pi \sigma^2} \exp\left(-\frac{x^2 + y^2}{2\sigma^2} \right)~,
\end{equation}
where $\phi$ is the total flux of the source, $\sigma = f / \left(2 \sqrt{2 \log 2}\right)$ and $f$ is the Full Width at Half Maximum (FWHM). The set of parameters for this source is ${ \boldsymbol{\theta}}_C = (\phi, f)$.
\item A Gaussian elliptical source
\begin{equation}
\varphi_E(x,y) = \frac{\phi}{2 \pi \sigma_M \sigma_m} \exp\left(-\frac{x'^2}{2 \sigma_M^2} - \frac{y'^2}{2\sigma_m^2} \right)~,
\end{equation}
where $\phi$ is the total flux of the source, $\sigma_M = f_M / \left(2 \sqrt{2 \log 2}\right)$, $\sigma_m = f_m / \left(2 \sqrt{2 \log 2}\right)$, $f_M$ is the major FWHM and $f_m$ is the minor FWHM. Moreover,
\begin{equation}
x' = \cos(\alpha) x + \sin(\alpha) y
\end{equation}
\begin{equation}
y' = -\sin(\alpha) x + \cos(\alpha) y
\end{equation}
where $\alpha$ is the angle between the semi-major and the $x$-axis measured counterclockwise. The set of parameters for the elliptical source is ${ \boldsymbol{\theta}}_E = (\phi, f_M, f_m, \alpha)$ \footnote{We point out that we actually optimize the two auxiliary variables: $\varepsilon \in (0,1)$ and $f$, satisfying  $f_M = f \; (1-\varepsilon^2)^{-\frac{1}{4}}$ and $f_m = f \;  (1-\varepsilon^2)^{\frac{1}{4}}$.}.
\item A double Gaussian circular source
\begin{equation}
\begin{array}{ll}
\varphi_D(x,y) & =  \frac{\phi_1}{2 \pi \sigma_1^2} \exp\left(-\frac{(x - x_0)^2 + (y - y_0)^2}{2\sigma_1^2} \right) \\ 
& +   \frac{\phi_2}{2 \pi \sigma_2^2} \exp\left(-\frac{(x + x_0)^2 + (y + y_0)^2}{2\sigma_2^2} \right) ~,\\
\end{array}
\end{equation}
where $\phi_j$ is the total flux, $\sigma_j = f_j / \left(2 \sqrt{2 \log 2}\right)$, $f_j$ is the FWHM ($j=1,2$) and $(x_0, y_0)$ are the coordinates of the center of one source (the other is symmetric with respect to the origin). The set of parameters for the double circular source is ${ \boldsymbol{\theta}}_D = (x_0, y_0, \phi_1, f_1, \phi_2, f_2)$.
\end{itemize}
The shapes of the three parametric sources are illustrated in Figure \ref{fig:fig-1}. 
In the next section we present the two computational methods for estimating the set of source parameters; without loss of generality, we will denote such set as ${ \boldsymbol{\theta}}$, which is intended to be either ${ \boldsymbol{\theta}}_C$, ${ \boldsymbol{\theta}}_D$ or ${ \boldsymbol{\theta}}_E$, according to the estimated shape.

\section{The approach to image reconstruction}

The multivariate optimization problem in hard X-ray solar imaging is typically represented by the minimum problem 
\begin{equation}\label{min_prob}
\argmin_{{ \boldsymbol{\theta}}\in {\cal D}} \hskip 0.2cm \chi^2({ \boldsymbol{\theta}}), 
\end{equation}
where ${\cal D}$ is the so-called feasibility region for the imaging parameter ${ \boldsymbol{\theta}}$ and $\chi^2({ \boldsymbol{\theta}})$ measures the square of the discrepancy between the experimental visibility amplitudes and the ones predicted by computing the Fourier transform of the source shape parameterized by $\boldsymbol{\theta}$. 
The solution of this problem usually relies on the {{RHESSI}} legacy \citep{2002SoPh..210....3L} and, specifically, on the vis$\_$fwdfit routine that computes the imaging parameters by means of a simplex approach. However, the number of visibilities provided by STIX is significantly smaller than the one provided by {{RHESSI}} and, at this stage of the calibration process, the information contained in STIX observations is even more limited. This is the reason why, in this first and preliminary study devoted to STIX imaging, we have implemented two more sophisticated approaches that have been explicitly designed in order to avoid local minima in the optimization process. 

\begin{figure}[h!]
    \centering
    \includegraphics[scale=0.4]{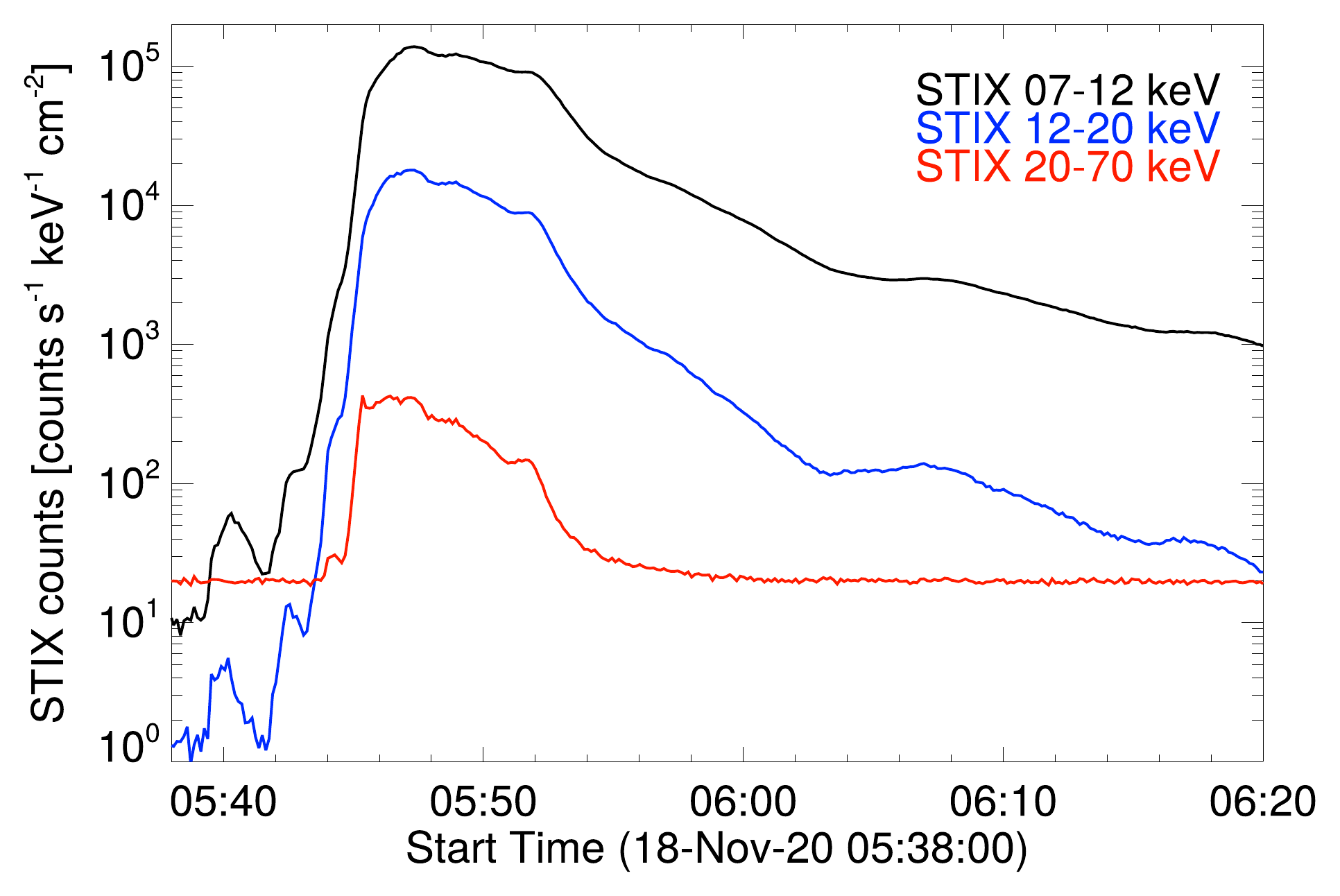}
    \caption{{{STIX}} light curves from November 18, 2020, 05:38 UT to November 18, 2020, 06:20 UT with different energy channels.}
    \label{fig:lightcurves}
\end{figure}

\begin{figure*}[ht!]
\centering

\centering
  \includegraphics*[height = .67\textwidth, angle=90, bb = 0 80 565 770]{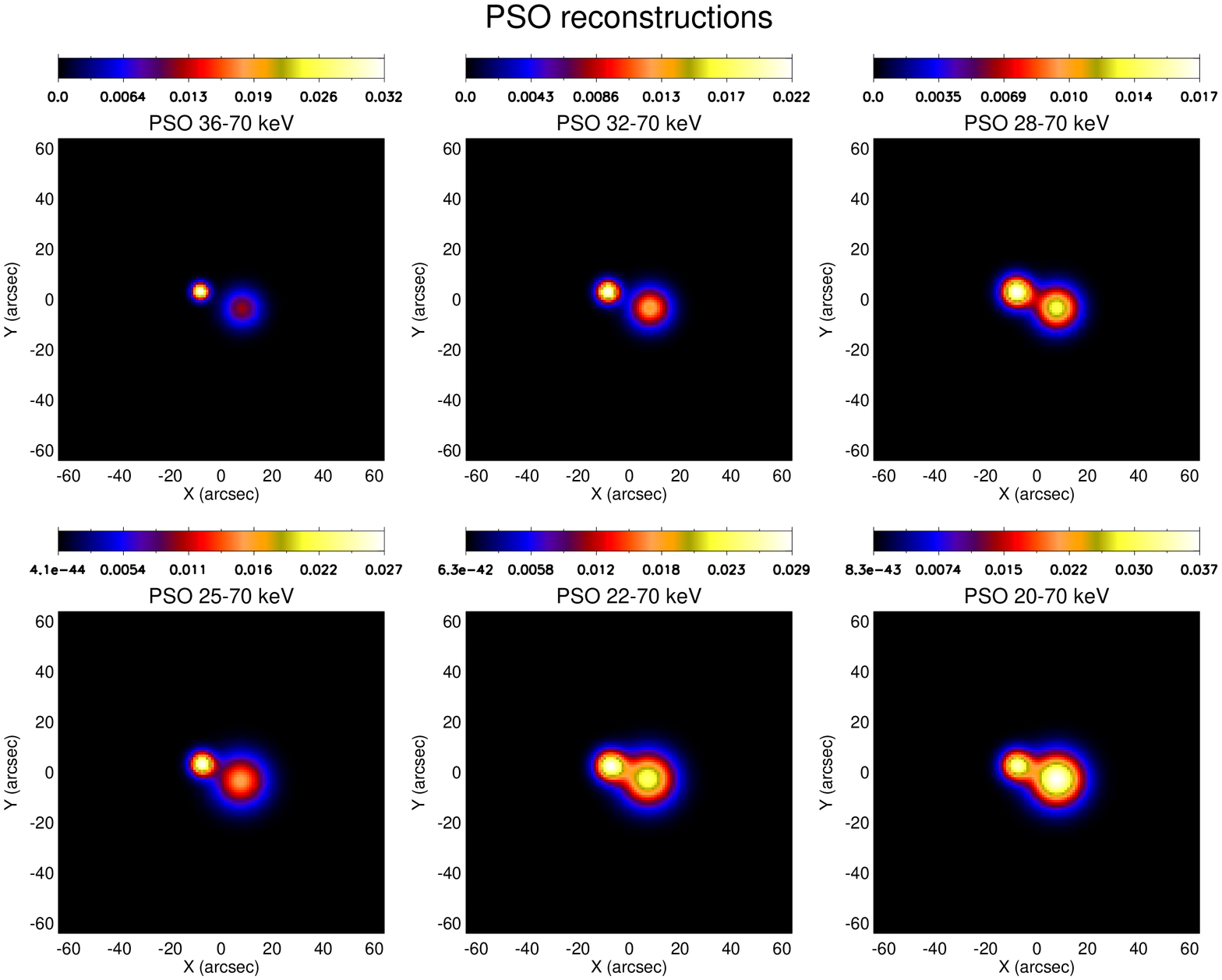}
  
   \vskip 0.3cm

\centering
    \includegraphics*[height = .67\textwidth, angle=90, bb = 0 80 565 770]{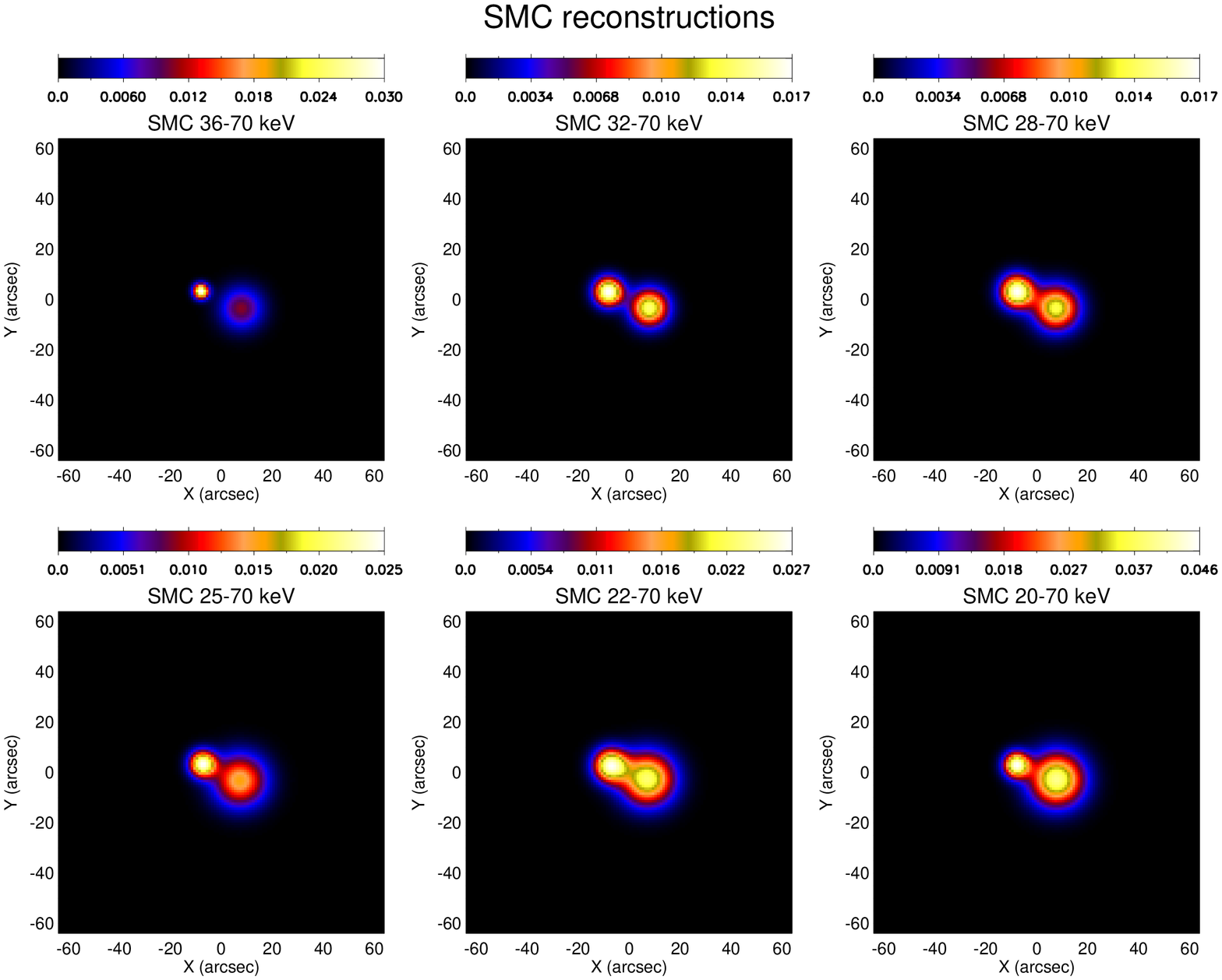}
    \caption{For each box, from top to bottom, left to right: the reconstructions computed via PSO (first box) and SMC (second box) from the visibility amplitudes observed by {{STIX}} at  05:45:30 UT -- 05:46:15 UT on the November 18, 2020,  corresponding to the energy channels $E_1$, $E_2$, $E_3$, $E_4$, $E_5$ and $E_6$, respectively. The SMC reconstructions use the conditional mean. }
    \label{figure:fig-3}
\end{figure*}

\begin{figure*}[ht!]
\centering
\includegraphics[height = .7\textwidth, angle =90]{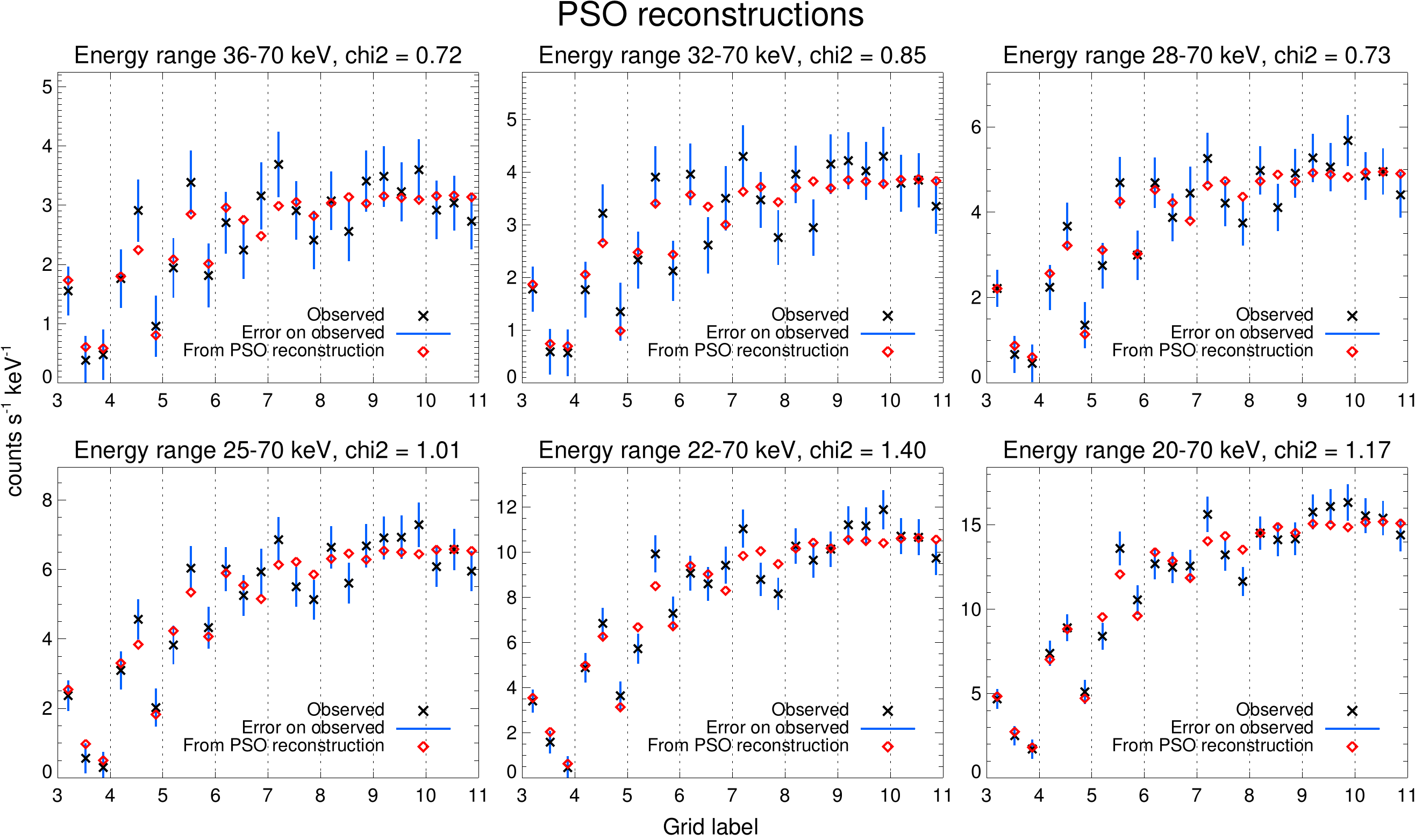} 
\\
\vskip 0.3cm
\centering
\includegraphics[height = .7\textwidth, angle = 90]{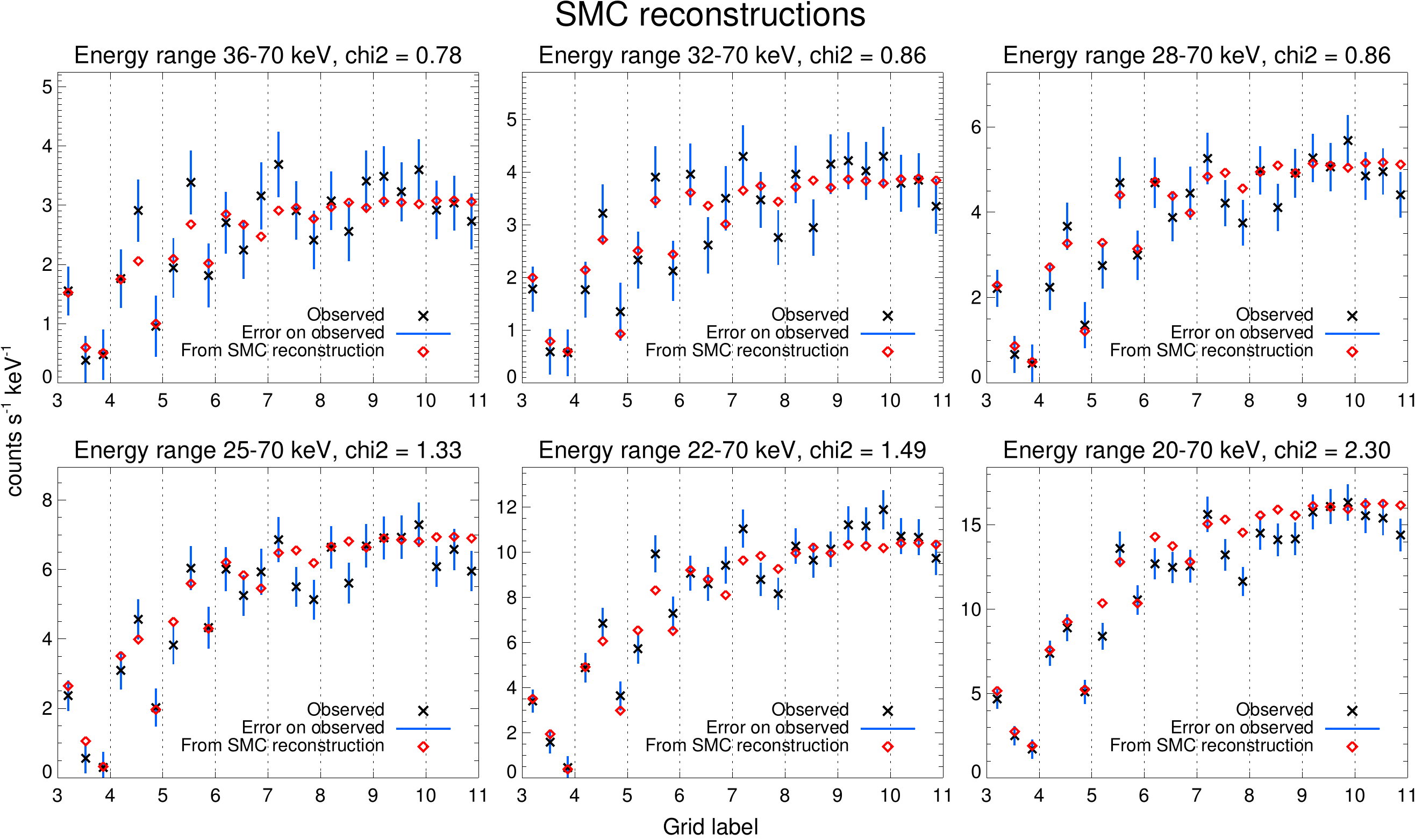}
\caption{For each box, from top to bottom, left to right: the amplitudes fits computed via PSO (first box) and SMC (second box) from the visibility amplitudes observed by STIX at  05:45:30 UT -- 05:46:15 UT on the November 18, 2020, corresponding to the energy channels $E_1$, $E_2$, $E_3$, $E_4$, $E_5$ and $E_6$, respectively. }
\label{figure:fig-4}
\end{figure*}

\subsection{Particle Swarm Optimization} Particle Swarm Optimization (PSO) \citep{eberhart1995particle,Qasem} does not require any strong assumptions on the issue defined in (\ref{min_prob}), and takes advantage of being usually more robust than deterministic strategies \citep[see e.g.,][]{Nocedal} in presence of multiple local minima. To briefly review it, let us consider a group of particles or birds which are represented as points in  the space  ${\cal D}$. The goal of swarm intelligence is to model the   trajectories of each single particle. Indeed, the target of a flock is to look for the maximum availability of food, i.e. the minimum of the objective function $\chi^2$. The trajectory of each single bird is updated at each step of the algorithm by taking into account both its \emph{selfish} and \emph{social  behavior}. Precisely, supposing that a particle visits some local minima then   the other birds can either: 
\begin{itemize}
	\item[i.] Move away from the flock towards the local minimum  (selfish behavior), or 
	\item[ii.] Stay close to the flock (social behavior).
\end{itemize}
Assuming that the two behaviors are well-balanced, then the flock gradually changes its position until the global minimum is reached. 

The {\tt IDL} implementation was designed following the guidelines given by \citet{Coello} and \citet{MATLAB}. Specifically, given a first random initialization, at the $j$-th iteration, each $i$-th bird position ${ \boldsymbol{\theta}}^{(j)}_i$ is defined via  its velocity  $\boldsymbol{w}^{(j)}_i$ and its best local position $\boldsymbol{l}^{(j)}_{i}$. Letting  $\boldsymbol{g}^{(j)}$ the global best position, the position of each particle is updated as  
\begin{equation}
{ \boldsymbol{\theta}}^{(j)}_i= \boldsymbol{\theta}^{(j-1)}_i+
 \boldsymbol{w}^{(j)}_i, 
\end{equation}
where
\begin{equation}
\begin{array}{ll}
\boldsymbol{w}^{(j)}_i= \omega^{(j)} \boldsymbol{w}^{(j-1)}_i & + 
\eta \; \boldsymbol{\psi}^{(j)}_i \odot \left( \boldsymbol{l}^{(j-1)}_i-\boldsymbol{\theta}^{(j-1)}_i\right)\\
& + \rho \; \boldsymbol{\xi}^{(j)}_i \odot \left(\boldsymbol{g}^{(j-1)}-\boldsymbol{\theta}^{(j-1)}_i\right), \\
\end{array}
\label{up_vel}
\end{equation}
for $i=1, \ldots, M$, where $\boldsymbol{\psi}^{(j)}_i$, $\boldsymbol{\xi}^{(j)}_i$ are randomly fixed and $\odot$ denotes the component-wise product.   The parameters $\eta$ and $\rho$ are the acceleration coefficients and $\omega^{(j)}$  is the inertia weight that is adaptively updated in our implementation \citep{Coello}.

Uncertainty quantification with PSO is determined by means of a confidence strip approach: several realizations of the input data are computed by randomly perturbing the experimental set of visibility amplitudes with Gaussian noise whose standard deviation is set equal to the errors on the measurements; for each realization PSO is applied; and, finally, the standard deviation of each optimized source parameter is computed.

\subsection{Sequential Monte Carlo}
The second approach to image reconstruction from {{STIX}} visibility amplitudes is based on a Bayesian source identification method \citep{sciacchitano2019sparse} which has proven capable of assessing, on RHESSI data \citep{sciacchitano2018identification}: 1) the likely number of sources in the image, 2) the parameters characterizing each source, and 3) the associated uncertainties.

Given the current restriction on the calibration of the imaging system, and the consequent loss of information on the position of the flaring source, we restrict here the usage of the Sequential Monte Carlo (SMC) method to the identification of one single shape per image, which is one of the three described in Section 2 (Fig. 1). The resulting method is an alternative approach to the one outlined in (\ref{min_prob}), as we are here interested in providing \emph{probabilistic estimates} of the set of source-related parameters $\boldsymbol{\theta}$. In a Bayesian setting\footnote{In a Bayesian setting, all the variables of the problem are considered as random variables, with information on their values encoded as their probability distributions. In the following we will use lower-case letters to indicate both the random variables and their specific values in a given instance.}, we need to describe the posterior distribution for the flare source parameters conditioned on the set of observed visibility amplitudes $p(\boldsymbol{\theta}|\bf{A})$, which has the following form
\begin{equation}
    p(\boldsymbol{\theta}|{\bf A}) = \frac{p({\bf A}|\boldsymbol{\theta})p(\boldsymbol{\theta})}{p({\bf A})}.
\end{equation}
Here, $p({\bf A}|\boldsymbol{\theta})$ is the likelihood function, expressing the probability that the observed visibility amplitudes ${\bf A}$ are produced by an image with parameters $\boldsymbol{\theta}$; $p(\boldsymbol{\theta})$ is the prior probability on the set of parameters; and $p({\bf A})$ is a normalisation factor.

The likelihood function is assumed to be \citep{2002SoPh}
\begin{equation}
    p({\bf A}|\boldsymbol{\theta}) \propto \exp\left(-\sum_{i=1}^N \frac{({\bf A} - {\bf F}_a(\varphi))^2_i}{2\sigma^2_i}\right),
\end{equation}
where the standard deviations are known.  Statistical errors in the visibility data are directly inferred from photon statistics in the corresponding detectors.  An additional allowance of $5\%$ is added in quadrature to account for systematic errors.


The choice for the prior distributions on the parameters is done according to \cite{sciacchitano2018identification}: 
\begin{eqnarray}
f \sim \mathcal{U}\left(\left[0, \frac{2}{3}FOV\right]\right)\\
\phi \sim \mathcal{U}\left(\left[0, \max {\bf A}\right]\right)\\
(x_0, y_0) \sim \mathcal{U}\left(\left[ - \frac{FOV}{2},  \frac{FOV}{2}\right], \left[ - \frac{FOV}{2},  \frac{FOV}{2}\right]\right)\\
\alpha \sim \mathcal{U}\big(\left[0, 180^{\circ}\right]\big)\\
\varepsilon \sim \mathcal{U}\left(\left[0, 1\right]\right),
\end{eqnarray}
where $FOV$ indicates the field of view of the map.

As a consequence of these choices for prior and likelihood, the posterior probability density results to be an analytically intractable function on a high-dimensional space. Therefore, to compute it we use the SMC method \citep{sciacchitano2018identification}, which  produces a sample set that is approximately distributed according to the posterior, and can be used to make inference on the values of the various parameters. Once the algorithm converges, the  parameters  of the reconstructed map are computed by using the mean values of the posterior distributions (for the different parameters). The notable advantage of this algorithm is that it is able to realize uncertainty quantification in a very elegant way. In fact, the availability of the posterior distribution allows the straightforward computation of the variance as the second moment associated to each source parameter, hence with no need to perturb the input visibility amplitudes' bag. With respect to what done by \cite{sciacchitano2018identification}, in the present study we have fixed the number and the type of sources. This is the reason why we used the mean values of the posterior distribution without conditioning it on these two random variables.

\section{The November 18, 2020 flaring event}

On November 18, 2020, during its cruise phase, {{STIX}} observed a series of flaring events with main peak of the lightcurve at around 05:50 UT (see Figure \ref{fig:lightcurves}).

\begin{figure}[ht!]
\centering
\includegraphics*[width = .47\textwidth, bb = 10 140 600 800]{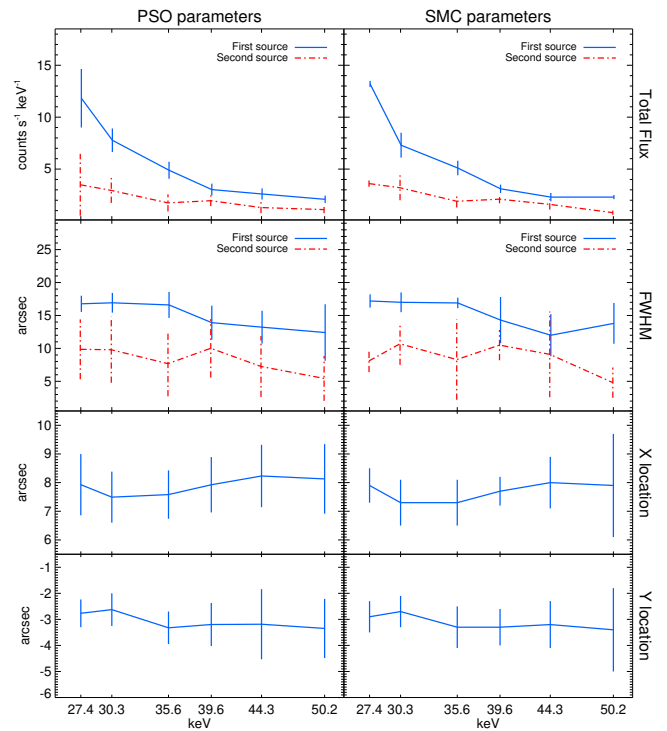}
\caption{Parameter values and related uncertainties reconstructed by PSO (left column) and SMC (right column) from the visibility amplitudes measured on November 18, 2020 from 05:45:30 to 05:46:15 UT in the energy channels $E_1$, $E_2$, $E_3$, $E_4$, $E_5$ and $E_6$, which are identified in the  the abscissa by their weighted mean energies. From top to bottom: total flux and FWHM of the two sources and $x$ and $y$ coordinates of the center of the first source (the other one is symmetric with respect to the origin). }
\label{fig:PSO_parameters}
\end{figure}

We used PSO and SMC to analyze the visibility amplitudes associated to this flare with the main aim to give a first and very preliminary assessment of {{STIX}} imaging performances. Therefore, the results illustrated in the next figures should be intended not as science products but as first hints of what {{STIX}} will allow doing for the investigation of hard X-ray flare physics when the instrument calibration process will be completed.

\subsection{The 05:45:30 UT -- 05:46:15 UT time window}

Focusing on the impulsive phase of the flare (05:45:30 UT -- 05:46:15 UT), we have applied PSO and SMC to the set of visibility amplitudes corresponding to six energy channels. Each bag was made of $24$ visibilities, $6$ visibilities being discarded as not yet well-calibrated (the discarded visibilities correspond to detectors with smallest pitch). The reconstruction results are presented in Figure \ref{figure:fig-3}, the corresponding fits in Figure \ref{figure:fig-4} and the parameter estimates in Figure \ref{fig:PSO_parameters}. The six energy channels have been selected by keeping the upper limit fixed at $70$ keV and gradually decreasing the lower limit of the channels. Precisely, we have taken $E_1$: 36--70 keV, $E_2$: 32--70 keV, $E_3$: 28--70 keV, $E_4$: 25--70 keV, $E_5$: 22--70 keV and $E_6$: 20--70 keV, which correspond to the weighted mean energies of $\bar{E}_1 = 50.2$ keV, $\bar{E}_2 = 44.3$ keV, $\bar{E}_3 = 39.6$ keV, $\bar{E}_4 = 35.6$ keV, $\bar{E}_5 = 30.3$ keV and $\bar{E}_6 = 27.4$ keV, respectively.  From the spectral analysis, the interval 36--70 keV should mainly contain the nonthermal emission and the corresponding reconstruction presents two distinguished rather compact sources, one at the bottom-right position (first source from now on) and the other one at the up-left position (second source from now on). While decreasing the lower limit of the energy channel, the flux associated to the first source significantly increases, while flux and dimension of the second source remain quite stable. This holds true for the reconstructions provided by both methods and, correspondingly, the small $\chi^2$ values confirm a notable statistical reliability of the results. A possible interpretation of this behavior is that the first source might be associated to the thermal emission with a nonthermal tail, while the second one reflects a nonthermal bremsstrahlung process. In any case, the interpretation of the physical nature of the different sources is not the goal of this paper, and the interpretation of the hard X-ray sources in this flare is left for when the full calibration of the STIX imaging system is available. 

\begin{figure}[h!]
\centering
   \fbox{\includegraphics[width=0.42\linewidth, height=0.34\linewidth]{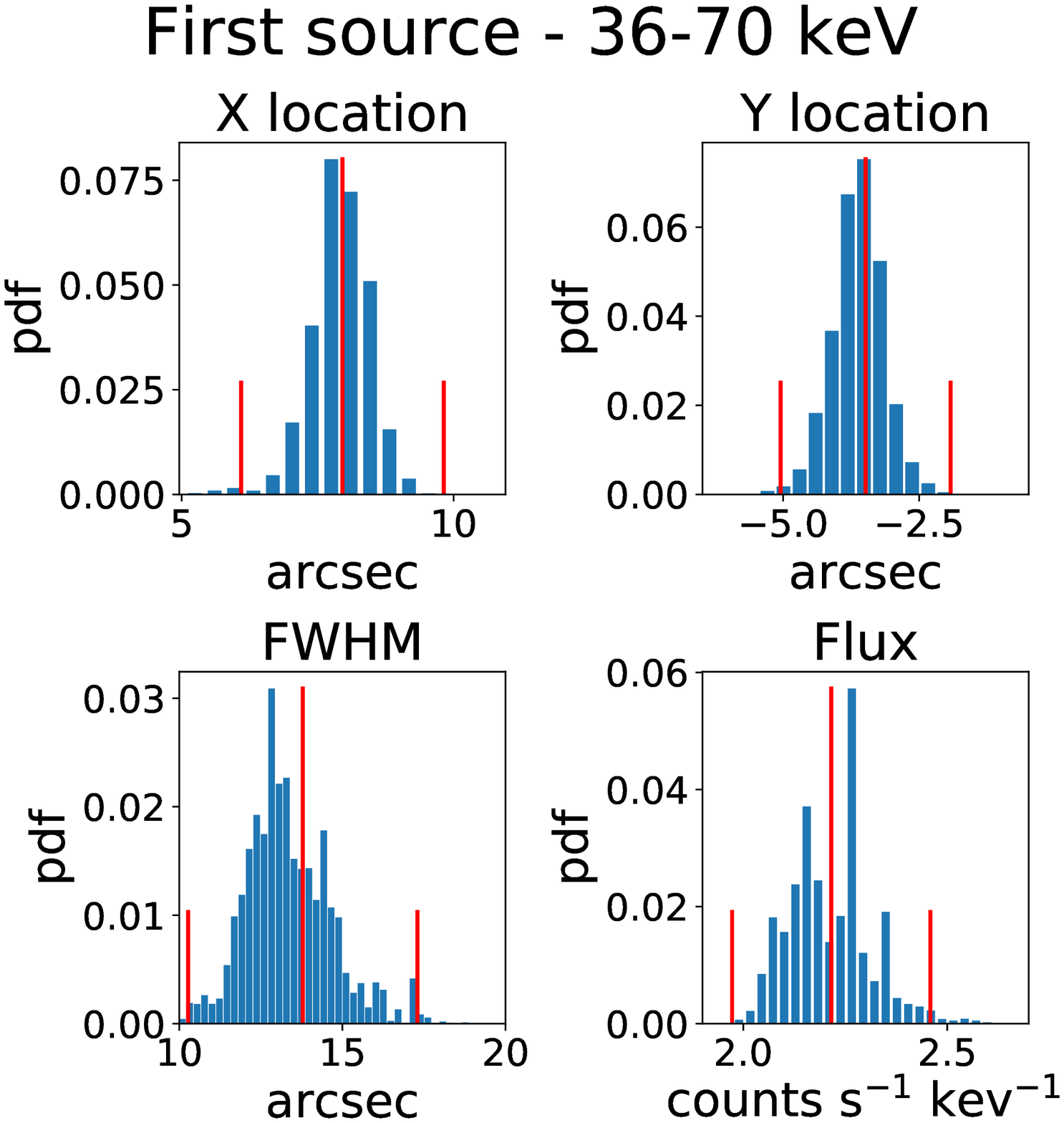}
   \includegraphics[width=0.42\linewidth, height=0.34\linewidth]{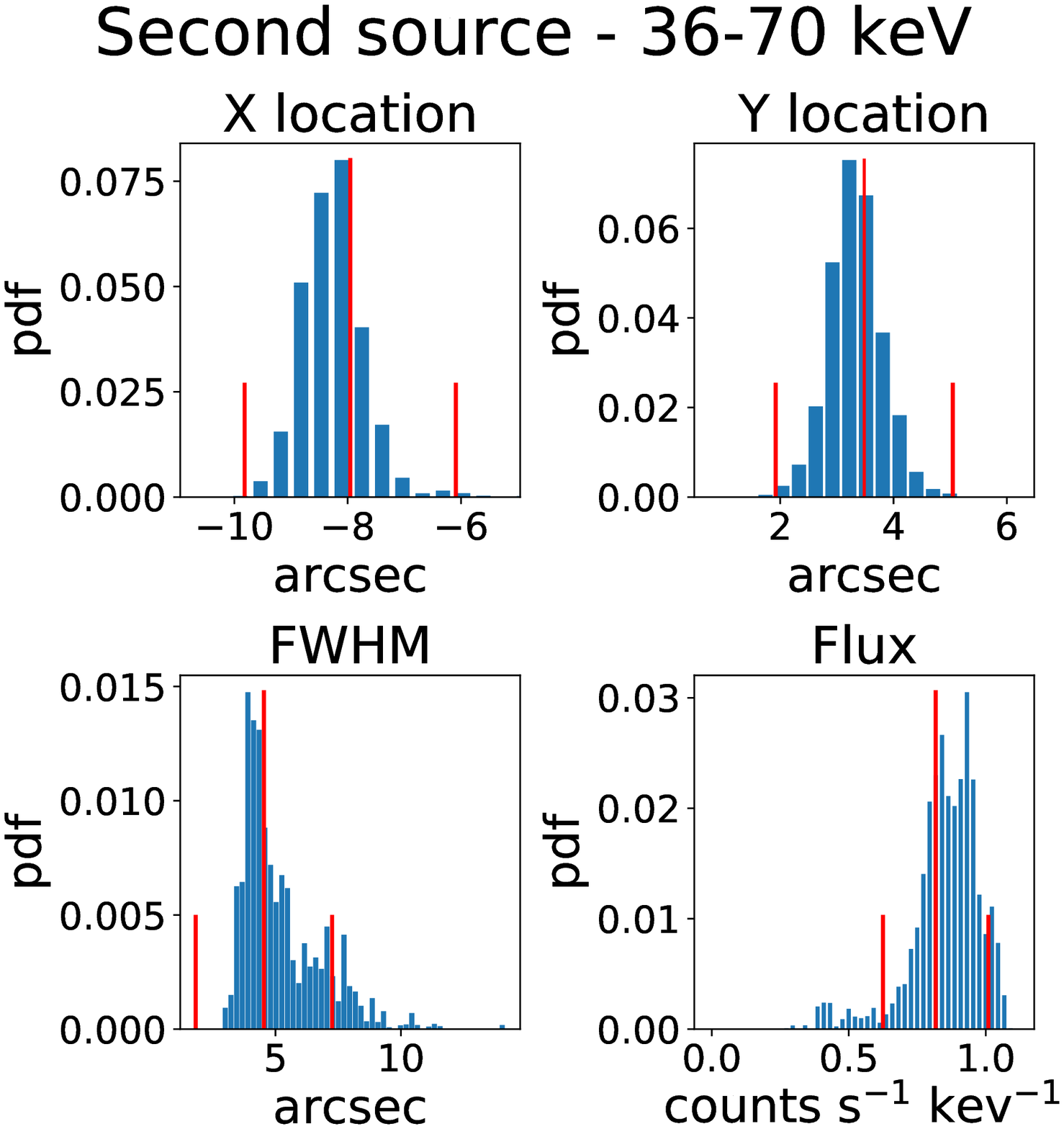}} 
   \fbox{\includegraphics[width=0.42\linewidth, height=0.34\linewidth]{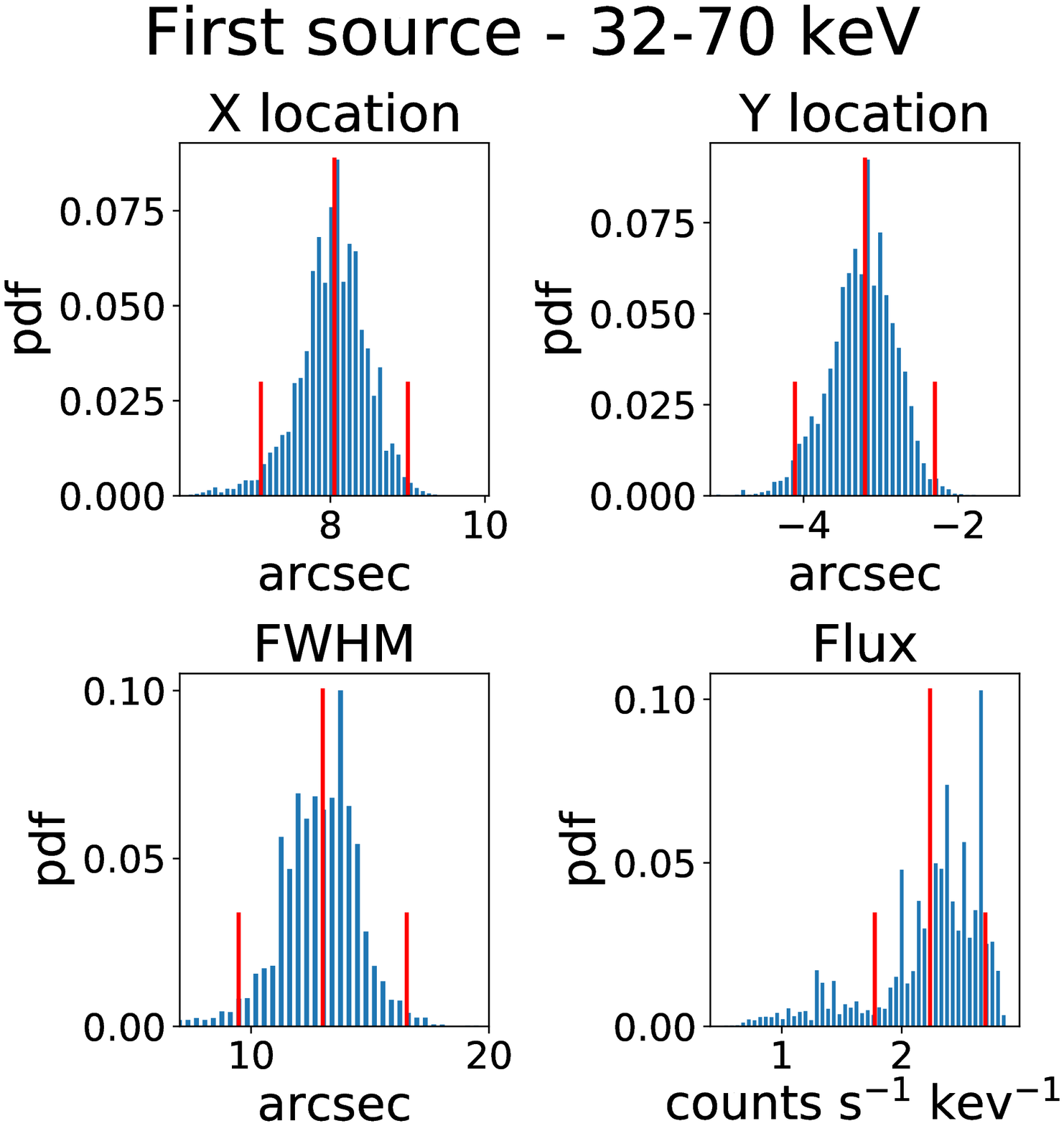}
   \includegraphics[width=0.42\linewidth, height=0.34\linewidth]{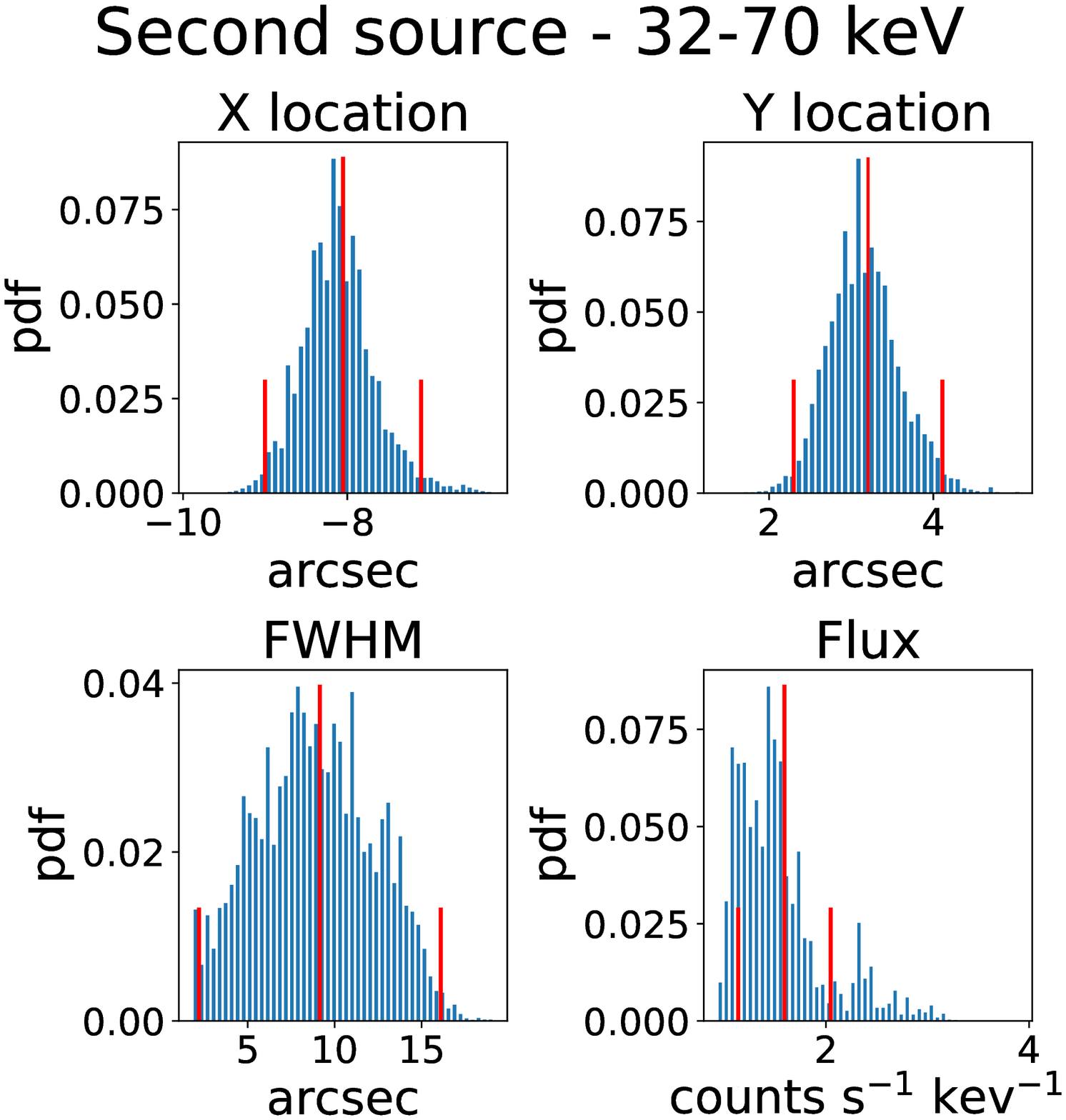}} 
   \\
   \vskip 0.1cm
   \fbox{\includegraphics[width=0.42\linewidth, height=0.34\linewidth]{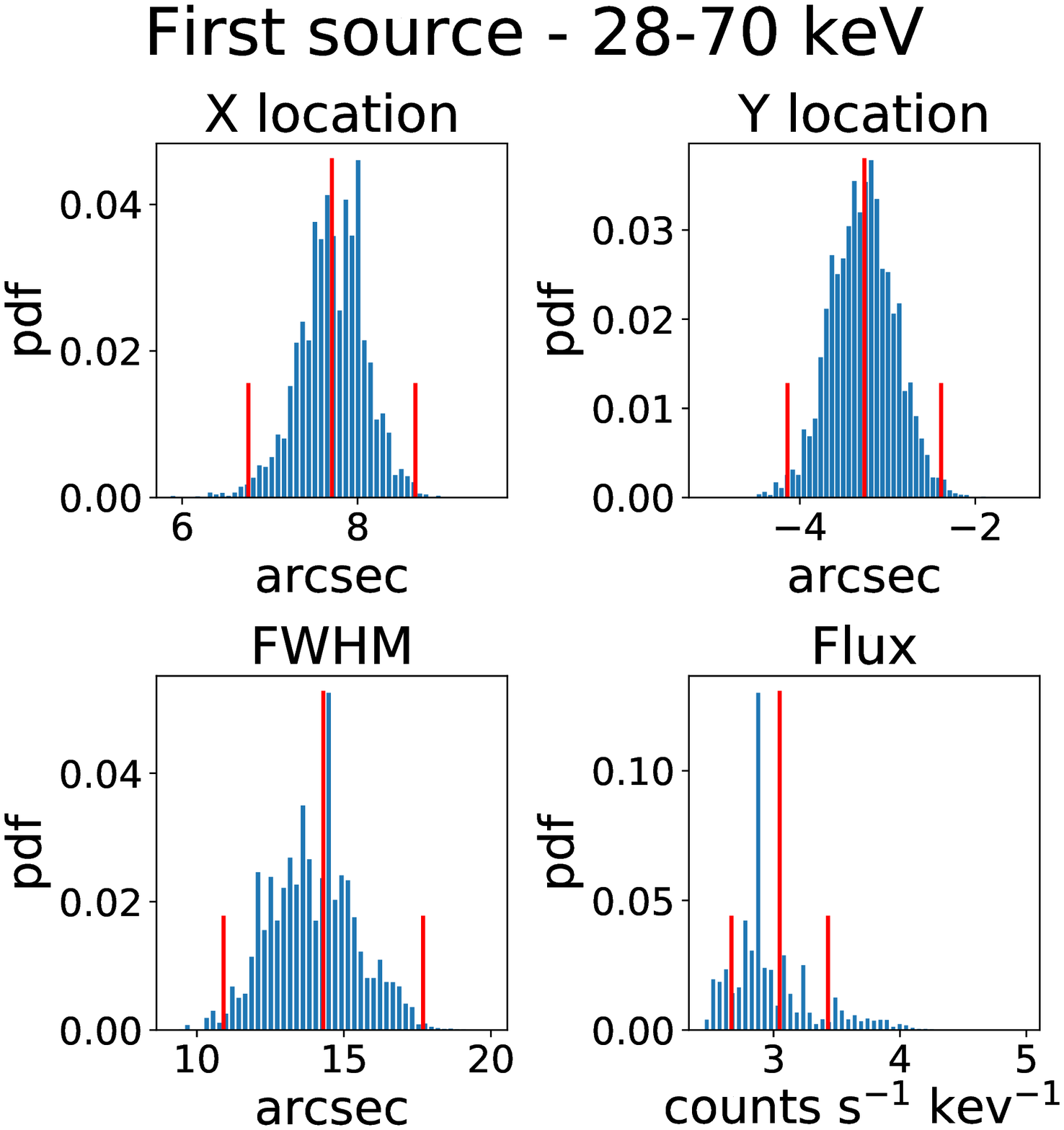}
   \includegraphics[width=0.42\linewidth, height=0.34\linewidth]{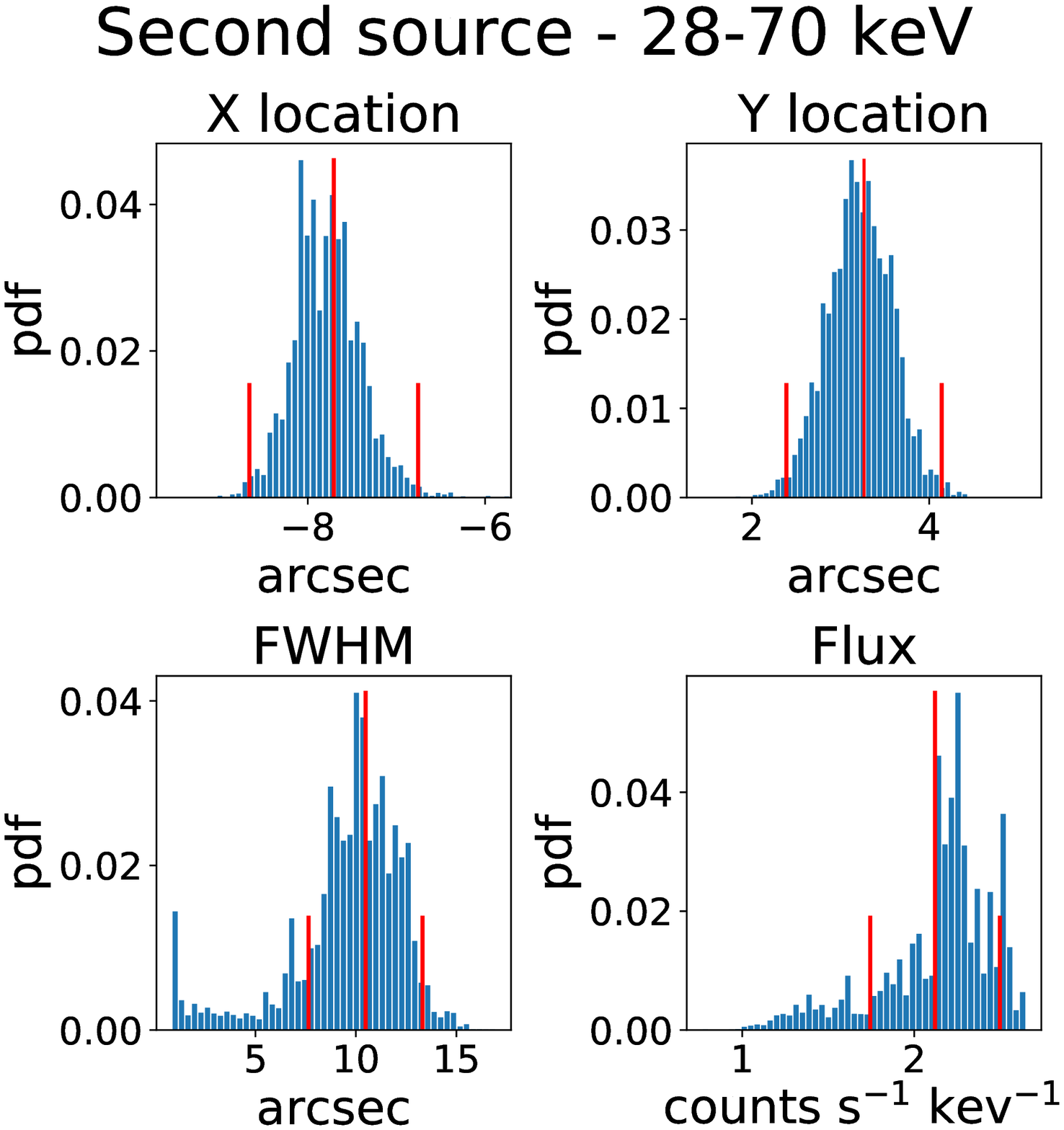}} 
   \fbox{\includegraphics[width=0.42\linewidth, height=0.34\linewidth]{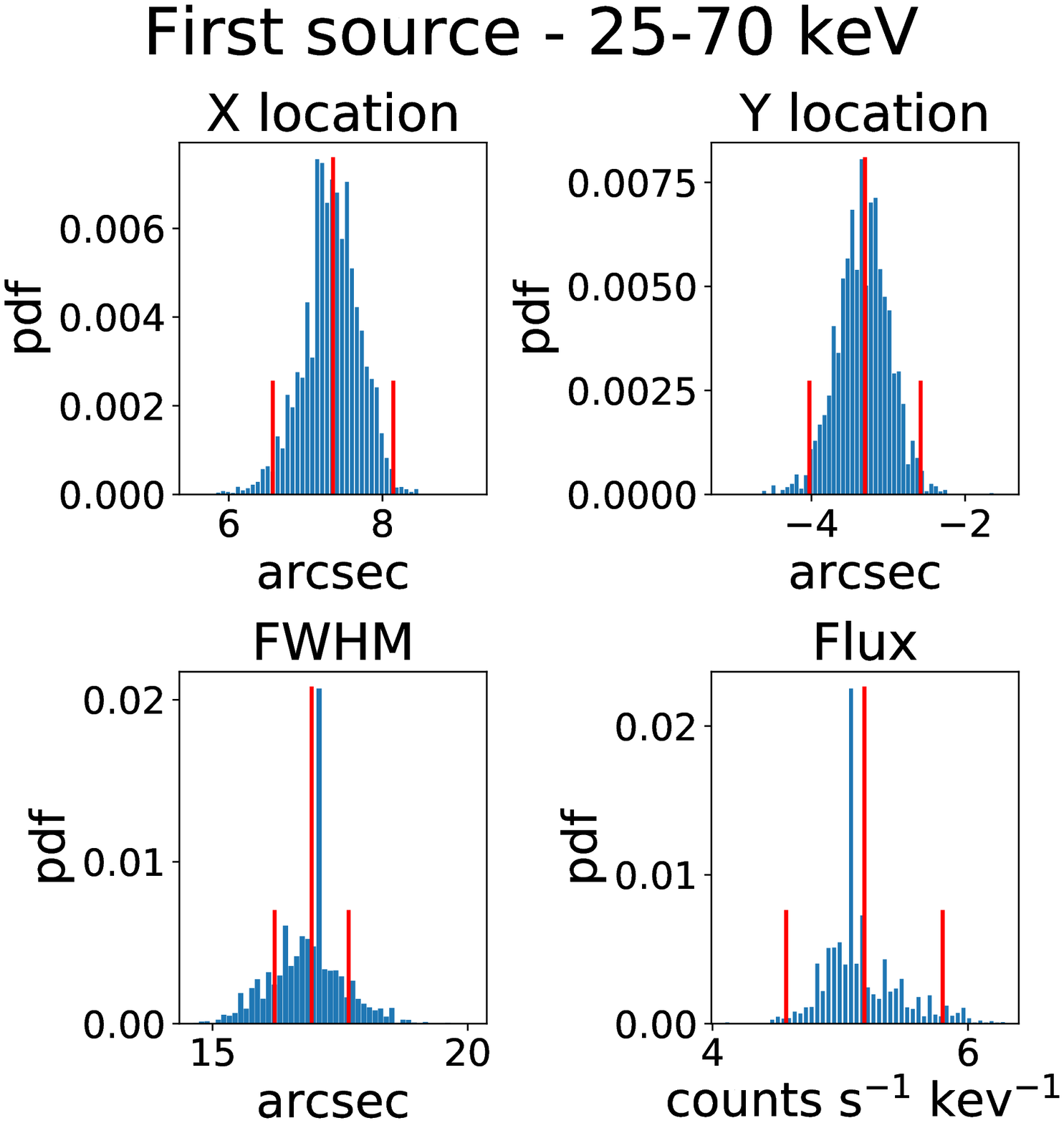}
   \includegraphics[width=0.42\linewidth, height=0.34\linewidth]{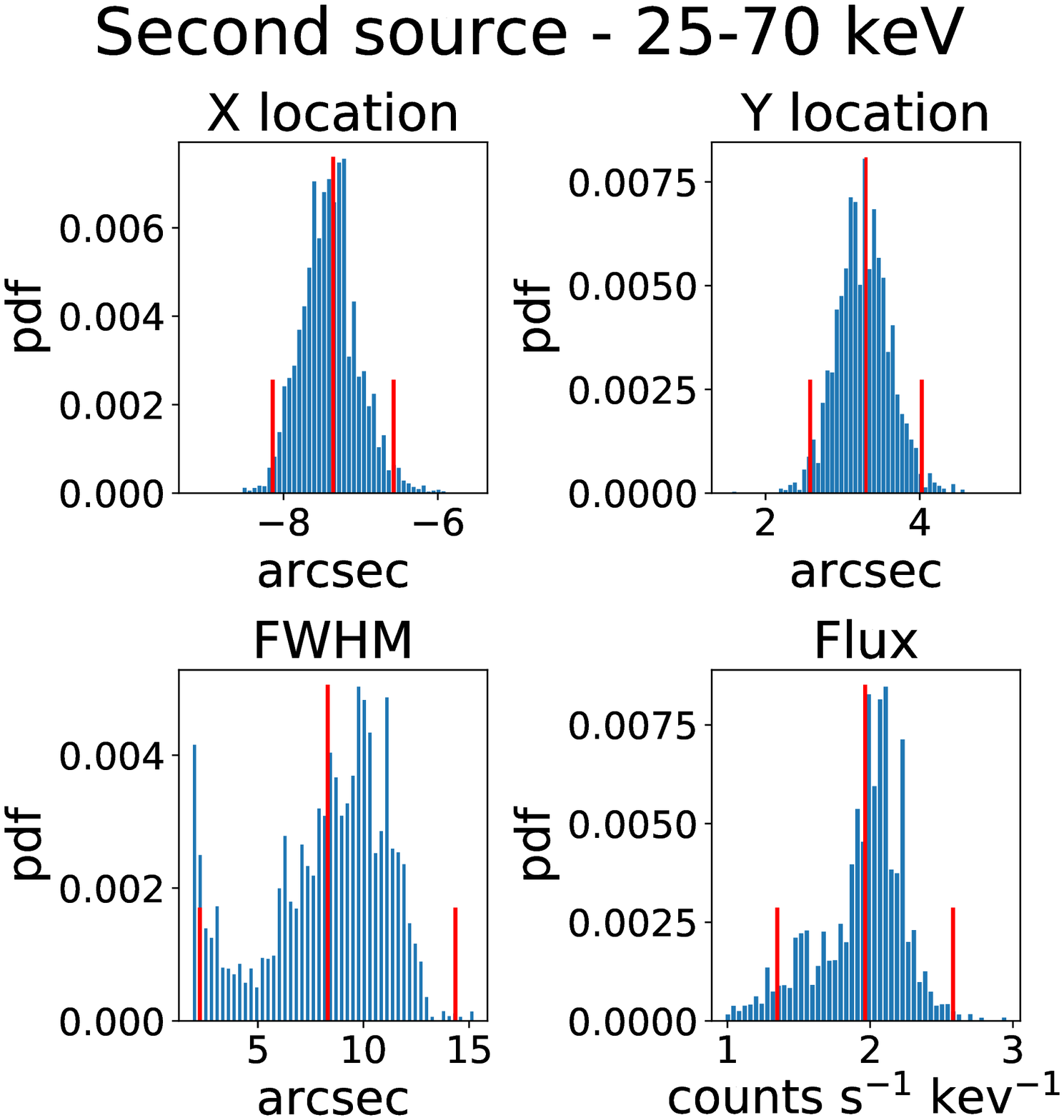}} 
   \\
   \vskip 0.1cm
   \fbox{\includegraphics[width=0.42\linewidth, height=0.34\linewidth]{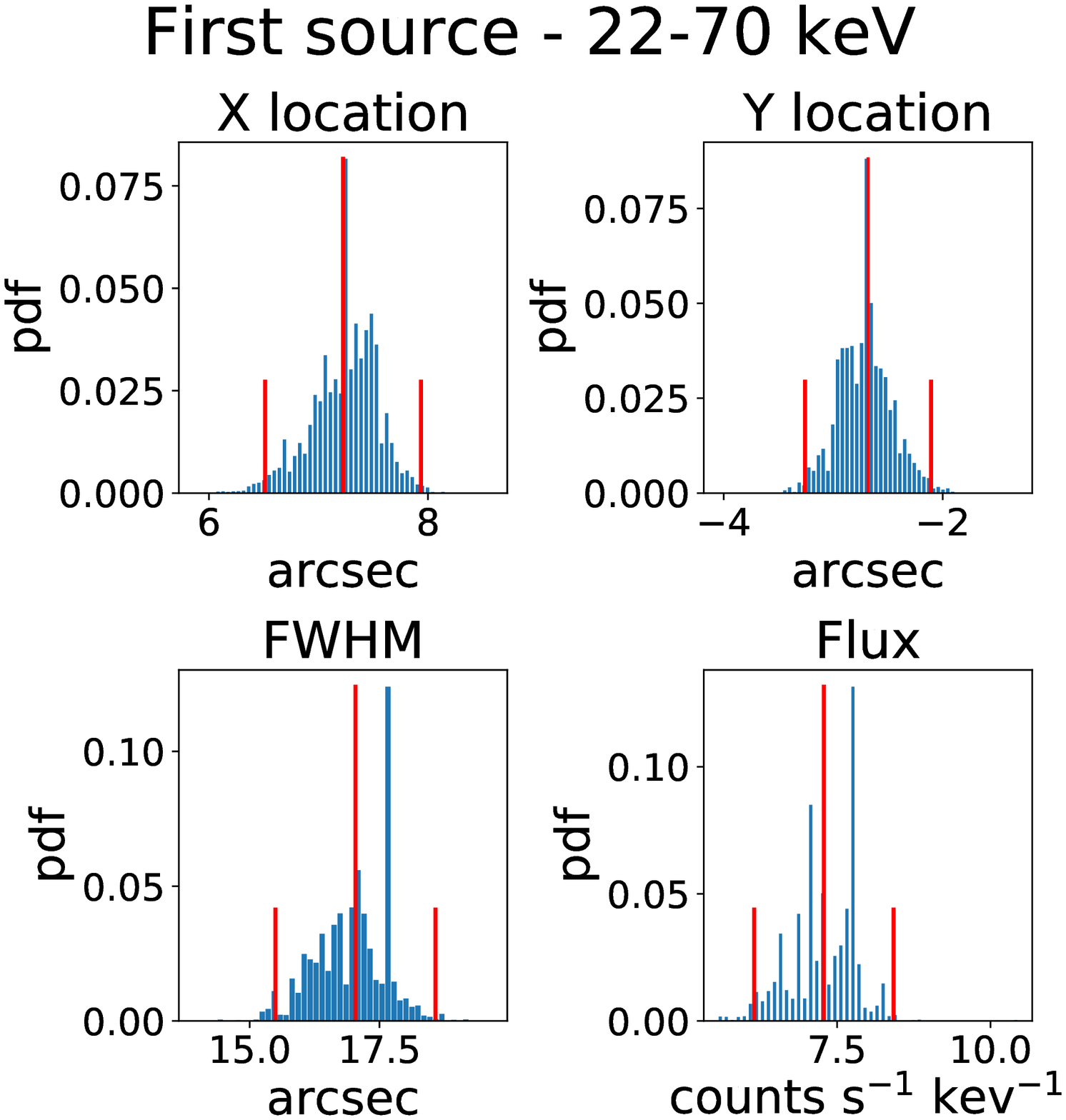}
   \includegraphics[width=0.42\linewidth, height=0.34\linewidth]{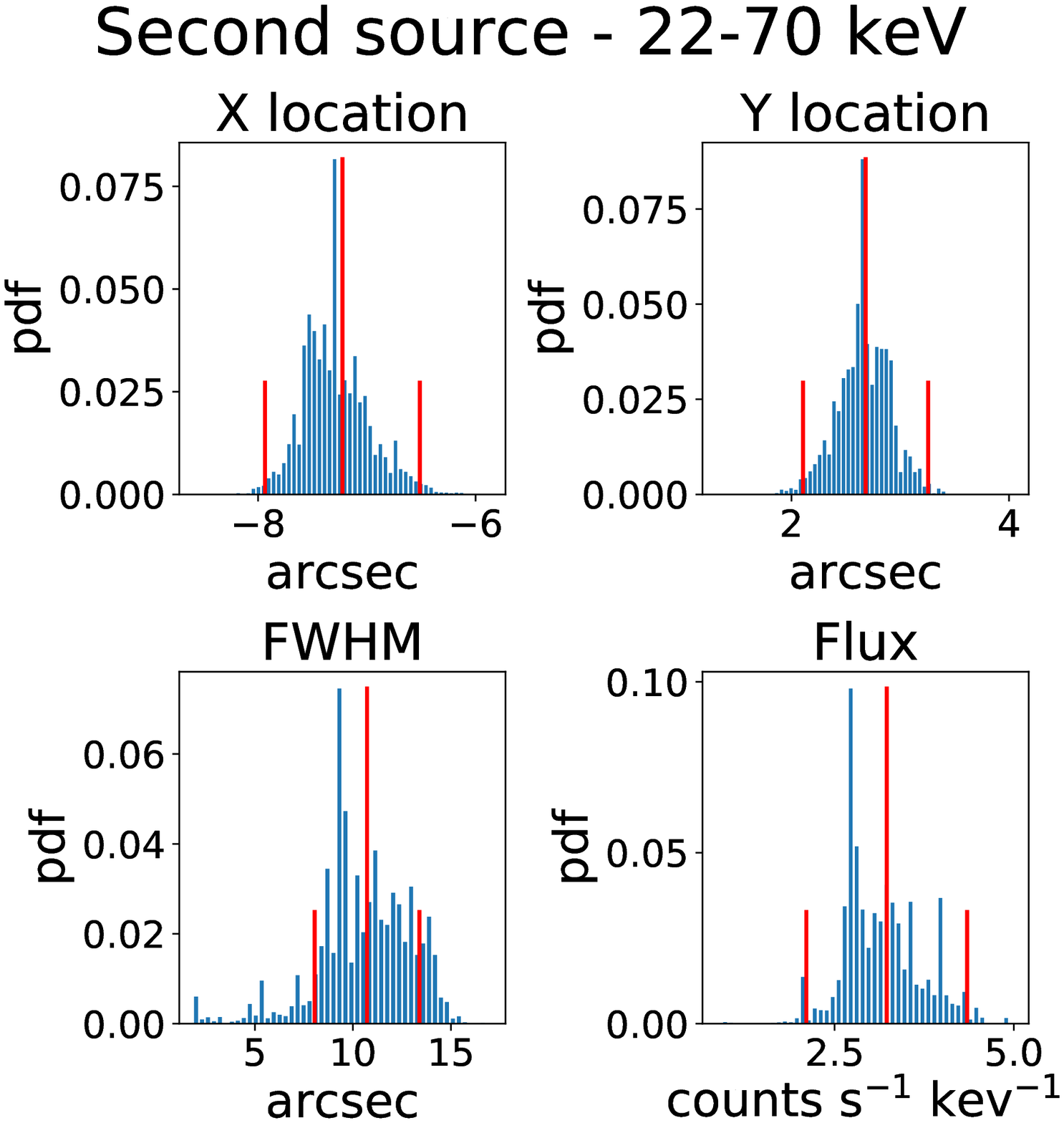}}       \fbox{\includegraphics[width=0.42\linewidth, height=0.34\linewidth]{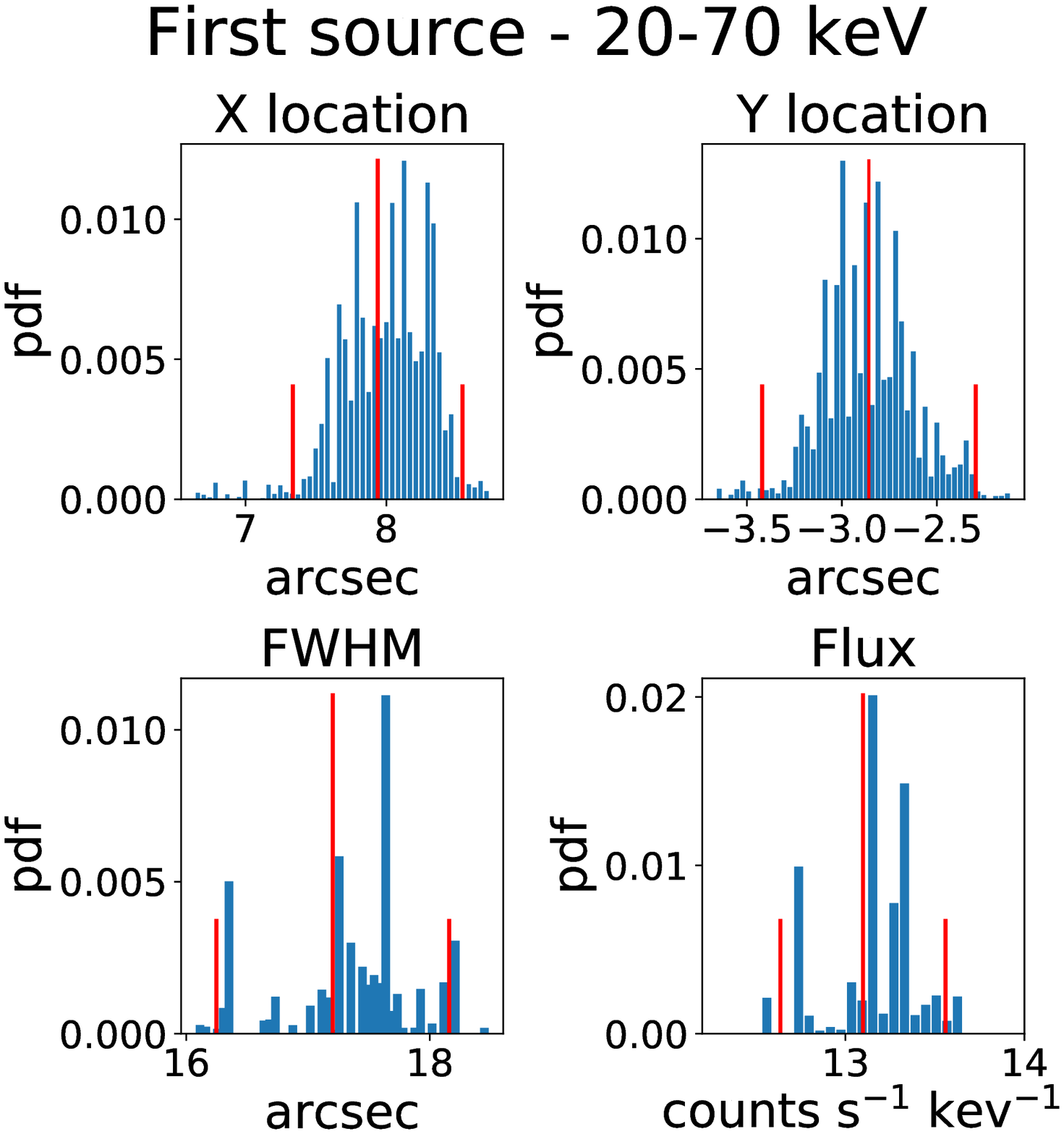}
   \includegraphics[width=0.42\linewidth, height=0.34\linewidth]{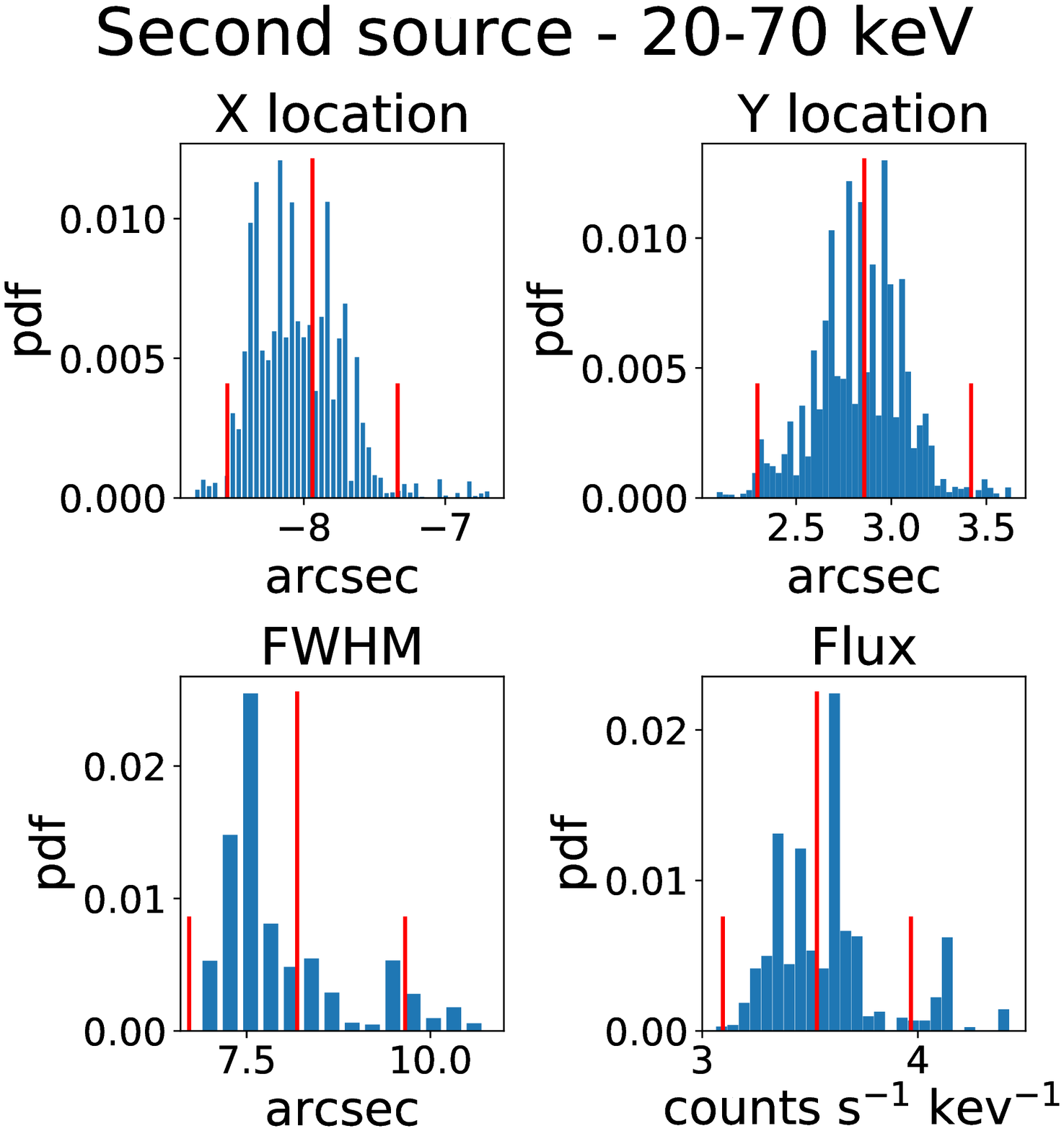}} \\
   \vskip 0.1cm
    \caption{From top to bottom, left to right, each box contains the parameter histograms returned by the SMC reconstruction from the visibility amplitudes observed by STIX at  05:45:30 UT -- 05:46:15 UT on the November 18, 2020, corresponding to the energy channels $E_1$, $E_2$, $E_3$, $E_4$, $E_5$ and $E_6$, respectively. 
    Each of the six boxes contains the probability distributions of the sources locations and associated FWHM and fluxes. In red, the estimated means (used to provide the reconstructed maps) and standard deviations.}
    \label{fig:hist_thermal_SMC}
\end{figure}

We note that, in the PSO analysis, uncertainty quantification has been performed by means of the confidence strip method, i.e. repeatedly perturbing the input data, re--running the algorithm for each data realization and computing the standard deviations over the set of reconstructed parameters.
On the contrary, the SMC method returns the estimated probability distributions of the parameters (via the histograms in Figure \ref{fig:hist_thermal_SMC}), and from them, means and standard deviations can be easily computed. This approach, which does not require the perturbation of the data, usually provides smaller uncertainties with respect to the ones provided by the confidence strip approach. The drawback of this method is an higher computational burden with respect to PSO. Indeed, each reconstruction took approximately 5 mins with SMC, while 1.5 mins with the PSO combined with confidence strip. We point out that reconstruction with PSO alone (i.e. not producing uncertainty estimation) is even faster, with $\sim$ 5 sec of computational time.

Another interesting aspect concerning the SMC histograms in Figure \ref{fig:hist_thermal_SMC} is that the width parameters, particularly the ones associated to the second source, are rather narrow. This is rather surprising, given that the six visibilities discarded from the analysis are the ones associated to the smallest angular resolution (smaller than $15^{\prime\prime}$). However, this super-resolution effect (which is present also in the case of the reconstructions provided by PSO) is a reasonable intrinsic consequence of the application of forward-fitting approaches: indeed, the use of predefined source shapes imposes strong constraints on the solution that may lead to an enhancement of the angular resolution of the reconstruction.
This behavior has been noticed also in the case of RHESSI visibilities, when a deterministic forward-fitting approach is applied for parameter estimation \citep[see point (5) in the Conclusion section of][]{2002SoPh..210..193A}.

Finally, some of the histograms in Figure \ref{fig:hist_thermal_SMC} present a shape that is far from Gaussian-like and is sometimes even close to a bi-modal one. This is most likely due to the fact that we are addressing a very ambiguous problem, in which two configurations characterized by interchanged sources provide the same $\chi^2$ value. Indeed, visibility amplitudes are not sensitive to parity transformations.

\subsection{Time resolution analysis}

\begin{figure*}[ht!]
\centering
    \includegraphics[height = 0.8 \textwidth, angle = 90]{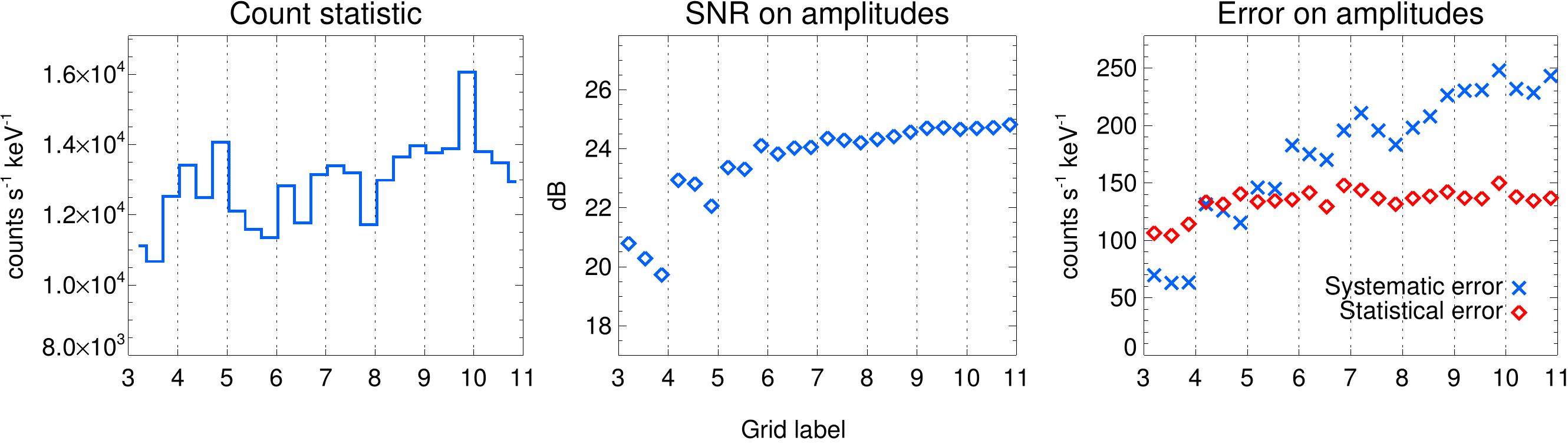} 
\caption{STIX observation for one second integration time, starting from 05:46:30 UT, in the energy channel 7–12 keV. Left panel: count statistic per detector. Middle panel: signal-to-noise ratio on visibility amplitudes. Right panel: contributions of the statistical and systematic errors on visibility amplitudes.}\label{figure:snr_1sec}
\end{figure*}

We considered the STIX observation at 05:46:30 UT of the November 18, 2020, with time integration of one second, in the energy range from $7$ to $12$ keV, in order to verify whether the count statistic, determined by the count rate recorded by the detectors and measured in counts keV$^{-1}$ s$^{-1}$,  was sufficient to allow the realization of reliable reconstructions. Figure \ref{figure:snr_1sec} shows that the count statistic is rather stable across detectors and that even at such a short integration time the statistical error is dominated by the $5\%$ systematic component.

\begin{figure*}[ht!]
\centering
    \includegraphics*[height = .6\textwidth, bb = 5 10 380 780, angle=90]{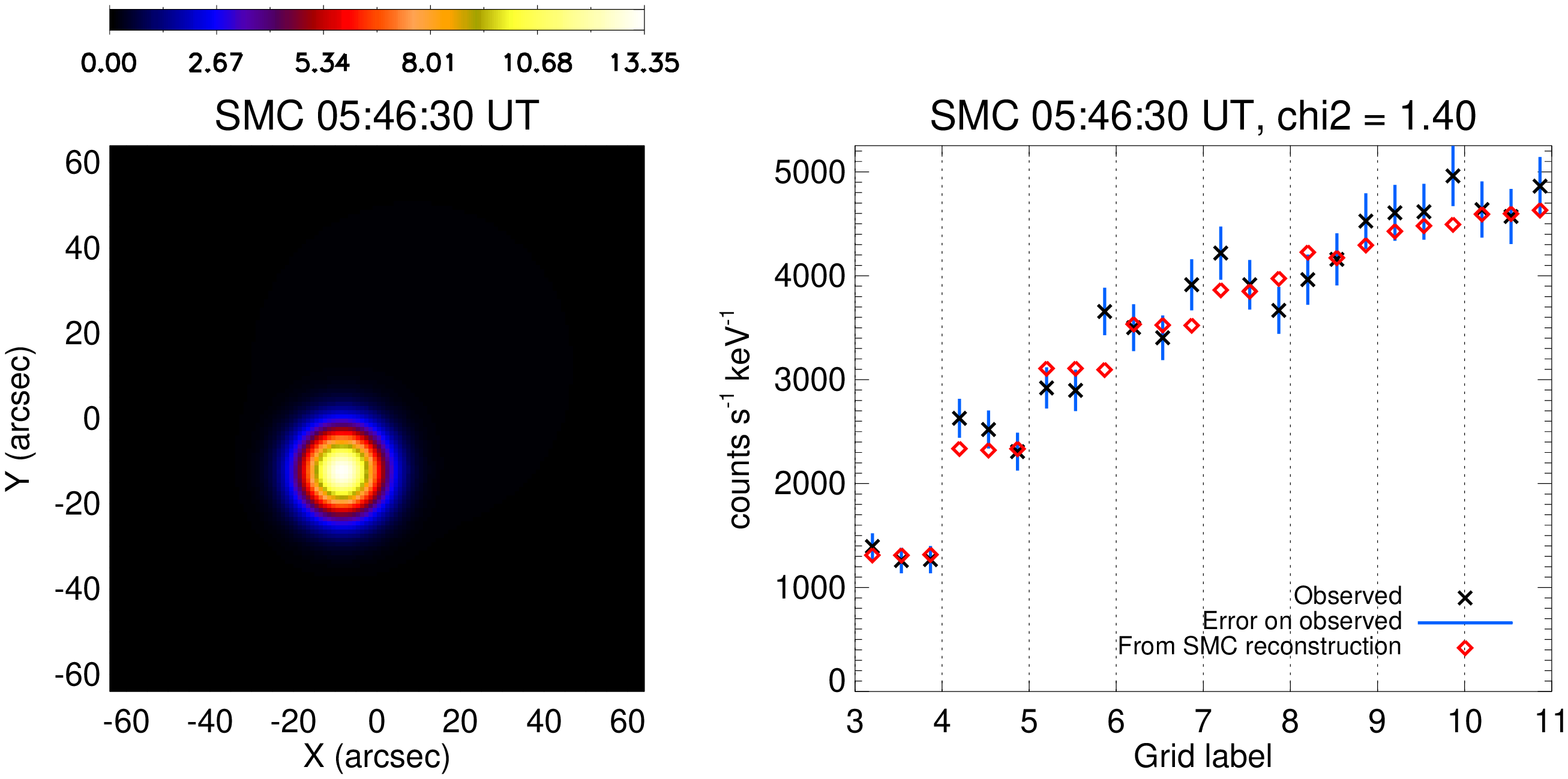}\\
    \vskip 0.4cm
    \includegraphics[scale=0.17]{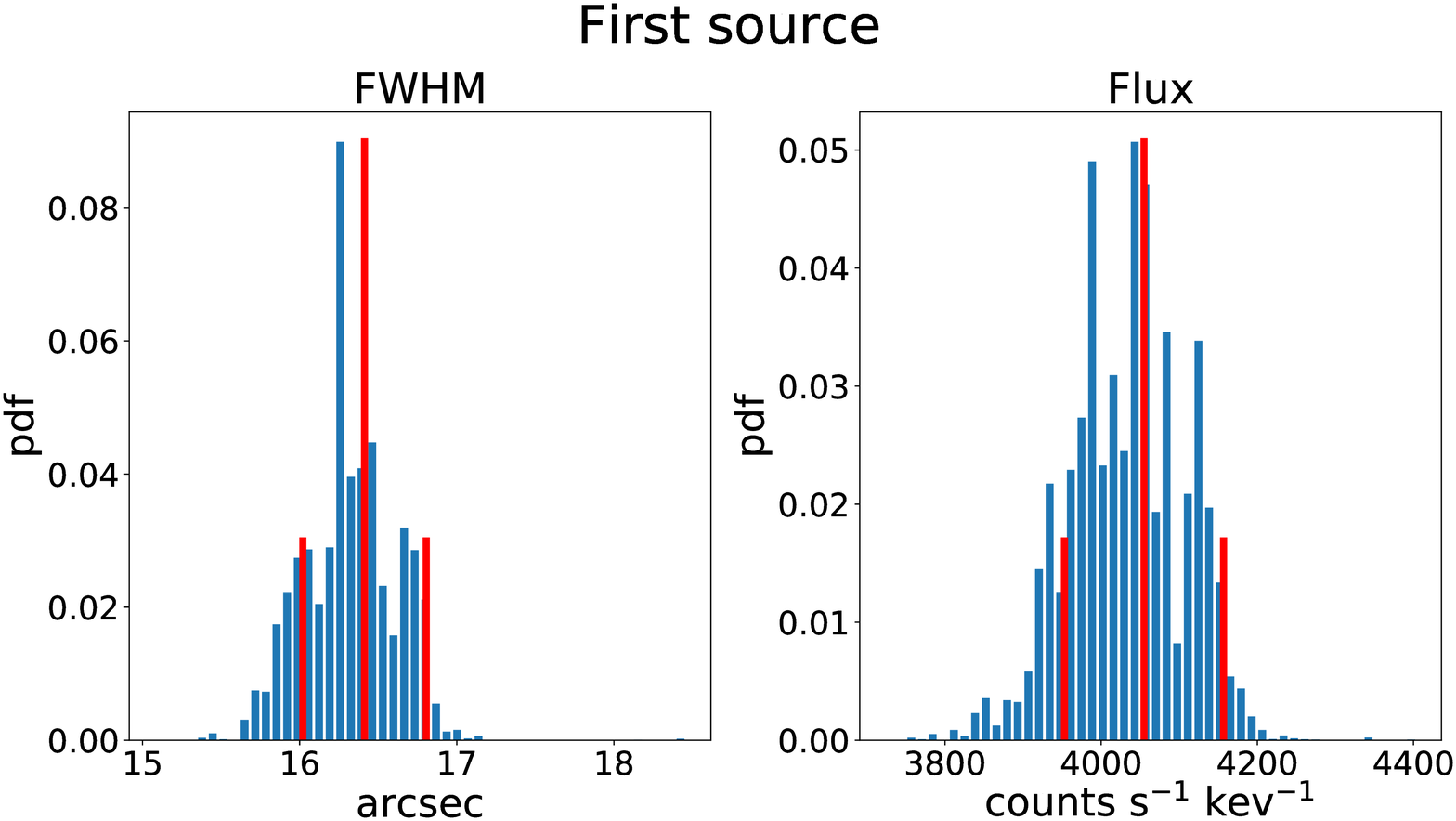} \hskip 0.2cm
    \includegraphics[scale=0.17]{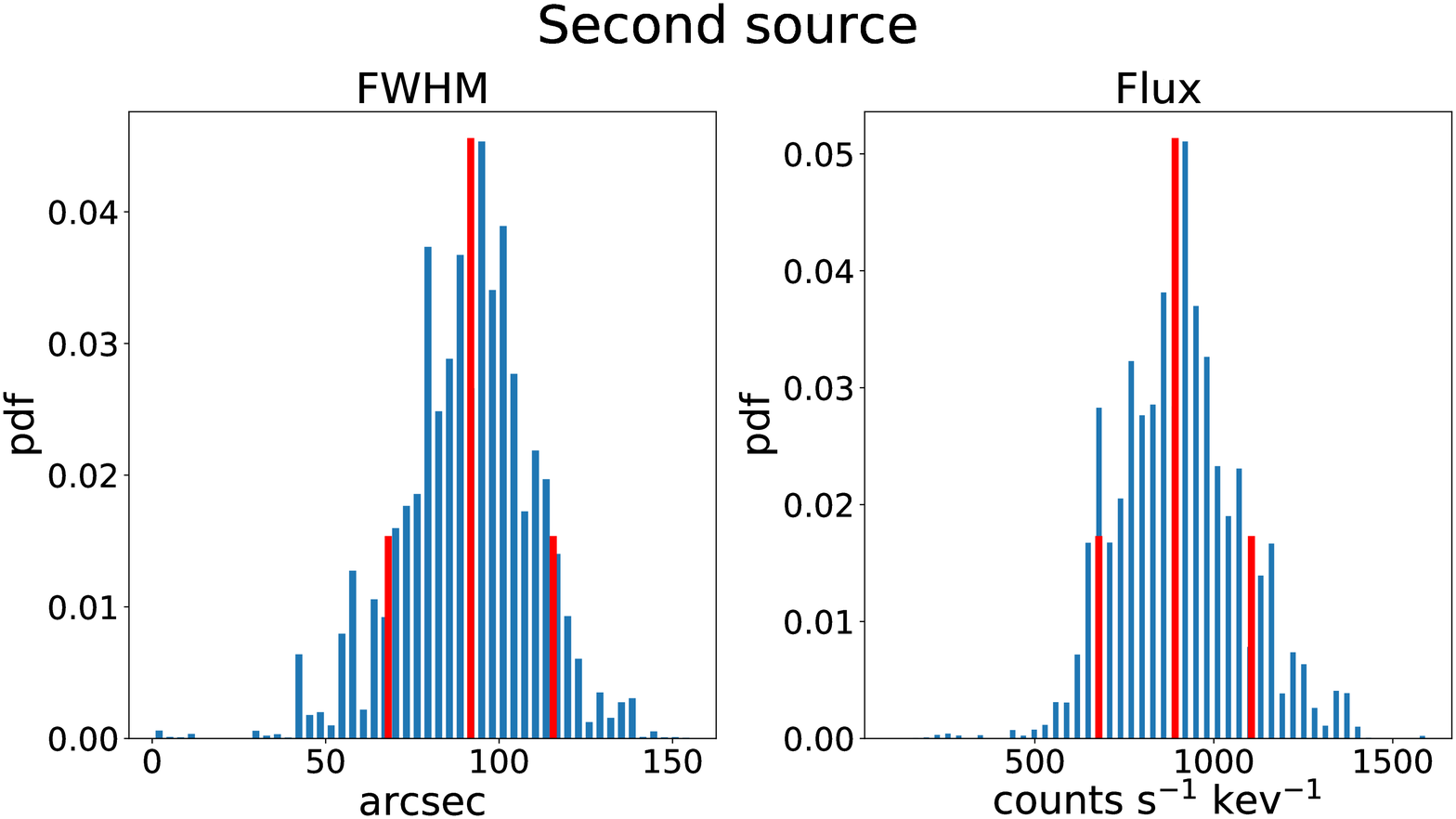} 
   \caption{SMC reconstruction from the visibility amplitudes observed by STIX at 05:46:30 UT on the November 18, 2020, with one second integration time in the energy channel 7--12 keV. Top left: flaring source modelled by means of a double-Gaussian input configuration; top right: corresponding data fit; bottom: histograms for the FWHM and flux probability distributions for the two sources. In red mean and standard deviations of the posterior distributions (used to give point estimates).}\label{figure:fig_1secSMC}
\end{figure*}

The application of SMC to the corresponding visibility amplitude bag produced the visualization and fitting in Fig.\,\ref{figure:fig_1secSMC}, top row, that correspond to an input configuration made of two Gaussian sources. We refer to \emph{first source} as the most energetic one, and \emph{second source} as the least energetic one. The histograms for the FWHM and flux parameters in the bottom row of that same figure show that the parameters for both sources are computed with sound uncertainty quantification. The values of such parameters are illustrated in Table \ref{tab:table_1secSMC}, where we also reported the results of the analysis performed by means of PSO. 

\begin{table*}[h!]
\caption{Estimated parameters and associated uncertainties computed by SMC and PSO for the visibility amplitudes recorded by STIX  at 05:46:30 UT on the November 18, 2020, with one second integration time in the energy channel 7--12 keV. }
\begin{center}
\begin{tabular}{ccccc}
\toprule
&\multicolumn{2}{c}{First source} &\multicolumn{2}{c}{Second source}\\
\cmidrule{2-5}
&FLUX (counts keV$^{-1}$ s$^{-1}$) & FWHM (arcsec)& FLUX (counts keV$^{-1}$ s$^{-1}$)  &FWHM (arcsec)\\
SMC  & 4055 $\pm$ 102 & 16.4 $\pm$ 0.4  & 892 $\pm$ 213  & 91.8 $\pm$ 23.8\\
PSO &4120 $\pm$ 129  &16.3 $\pm$ 0.5  &958 $\pm$ 295   &97.8 $\pm$ 17.2 \\
\bottomrule 
\end{tabular}
\end{center}
\label{tab:table_1secSMC}
\end{table*}

We note here that the integration time of the November 18, 2020 observation was set to $1$ second. This time interval is currently considered as the maximal time resolution achievable by STIX, although tests at faster time cadences (down to 0.1 second) will be carried out later in 2021. Considering our initial results at 1 second cadence, we are confident that sub-second time resolution imaging will be possible at least in the thermal range.

\section{Comments and conclusions}

This study represents a preliminary attempt to reconstruct hard X-ray images of solar flares from data collected by STIX during its cruise phase. As a consequence, these results should be considered with caution, first of all because the current calibration stage of the instrument does not allow the exploitation of the visibility phases, so that the non--linear image reconstruction problem of determining the flaring source from the visibility amplitudes is highly ambiguous.
It follows that the only approaches we could implement were the ones based on the forward fitting of very simple parametric source shapes and the only information we could try and determine were related to the flux and dimensions of such shapes and, in the case of a configuration made of two Gaussian sources, their relative position. 

However, even with these limitations, some hints can be deduced from the results of this analysis. For example, STIX seems to allow a high temporal resolution analysis, providing data with significantly high signal--to--noise ratio even in the case of very short integration times.
Further, a simple spectral analysis provided us with results that are consistent with a scenario in which a nonthermal footpoint persists at higher energies, while a (probably coronal) thermal component becomes more evident when lower energies are included in the processing. 

Finally, the two parametric imaging approaches considered in this work provide rather similar reconstructions. However, from a methodological viewpoint, the two methods have a key difference: PSO realizes uncertainty quantification through the confidence strip method, which is fast but does not leave the measurements untouched, possibly leading to suboptimal results. SMC provides a full probabilistic description of the reconstructed shapes, and therefore a more robust uncertainty estimation, but at the cost of an higher computational burden. We point out that both methods can be easily extended for use with fully calibrated visibilities (i.e., when even the visibility phases will be calibrated) and that they will be included in the STIX data analysis software of the mission.

\begin{acknowledgements}
{\em{Solar Orbiter}} is a space mission of international collaboration between ESA and NASA, operated by ESA. The STIX instrument is an international collaboration between Switzerland, Poland, France, Czech Republic, Germany, Austria, Ireland, and Italy. AFB is supported by the Swiss National Science Foundation Grant 200021L\_189180 for STIX. PM, EP, FB and MP acknowledge the financial contribution from the agreement ASI-INAF n.2018-16-HH.0. 
SG acknowledges the financial support from the "Accordo ASI/INAF Solar Orbiter: Supporto scientifico per la realizzazione degli strumenti Metis, SWA/DPU e STIX nelle Fasi D-E".

\end{acknowledgements}

\bibliographystyle{aa.bst}
\bibliography{bib_stix}

\end{document}